\documentclass[aps,prb,showpacs,twocolumn]{revtex4}

\usepackage[T1]{fontenc}
\usepackage{amssymb}
\usepackage{amsbsy}		
\usepackage{epsfig}

\begin{document}

\title{Antiferromagnetism and single-particle properties in the
  two-dimensional half-filled Hubbard model: a non-linear sigma model
  approach }  
\author{K. Borejsza and N. Dupuis}
\affiliation{ Laboratoire de Physique des Solides, CNRS UMR 8502, \\
  Universit\'e Paris-Sud, 91405 Orsay, France }
\date{December 4, 2003}


\begin{abstract}  
We describe a low-temperature approach to the two-dimensional
half-filled Hubbard model which allows us to study both
antiferromagnetism and single-particle properties. This approach
ignores amplitude fluctuations of the antiferromagnetic (AF) order
parameter and is valid below a crossover temperature $T_X$ which marks
the onset of AF short-range order. Directional fluctuations (spin
waves) are described by a non-linear sigma model (NL$\sigma$M) that we
derive from the Hubbard model. The parameters of the NL$\sigma$M, the spin
stiffness and the spin-wave velocity, are calculated as a function
of the Coulomb repulsion $U$. The NL$\sigma$M is solved by a saddle-point
approximation within the CP$^1$ representation where the N\'eel field
is parametrized by two Schwinger bosons. At zero temperature, there is
always Bose condensation of the Schwinger bosons, which signals AF
long-range order for any value of the Coulomb repulsion. At finite
temperature, the AF long-range order is suppressed (in agreement with
the Mermin-Wagner theorem), but the AF correlation length remains
exponentially large. In the CP$^1$ representation, the
fermion field is naturally expressed as the product of a Schwinger
boson and a pseudo-fermion whose spin is quantized along the
(fluctuating) N\'eel
field. This allows us to write the fermion Green's function as the
product (in direct space) of the Schwinger boson propagator (which is
derived from the NL$\sigma$M) and the pseudo-fermion
propagator. At zero temperature and weak coupling, our results are
typical of a Slater antiferromagnet. The AF gap is exponentially
small; there are well-defined Bogoliubov quasi-particles (QP's)
(carrying  most of the spectral weight) coexisting with a high-energy
incoherent excitation background. As $U$ increases, the Slater antiferromagnet
progressively becomes a Mott-Heisenberg antiferromagnet. The
Bogoliubov bands evolve into Mott-Hubbard bands separated by a large
AF gap. A significant fraction
of spectral weight is transferred from the Bogoliubov QP's to
incoherent excitations. At finite temperature, there is a
metal-insulator transition between a pseudogap phase at weak coupling and a 
Mott-Hubbard insulator at strong coupling. Finally, we point out that our
results straightforwardly translate to the half-filled attractive
Hubbard model, where the ${\bf q}=(\pi,\pi)$ charge 
and ${\bf q}=0$ pairing fluctuations combine to form an order
parameter with SO(3) symmetry.  
\end{abstract}


\maketitle

\section{Introduction} 
\label{sec0introduction}

The Hubbard model\cite{Hubbard63,Gutzwiller63,Kanamori63,Gebhard97}
and its generalizations play a key role in the 
description of strongly correlated fermion systems such as high-$T_c$
superconductors, heavy fermions systems, or organic
conductors. \cite{Imada98} Despite its simplicity (the model is
defined by two parameters, the inter-site hopping amplitude $t$ and the
local Coulomb interaction $U$, and the symmetry of the lattice), exact
solutions or well-controlled approximations exist only in a few
special cases like in one-dimension\cite{Voit94} (1D) or in the limit
of infinite dimension.\cite{Georges96}  

It is now well established that the ground state of the {\it
half-filled} Hubbard model on a cubic or square lattice has
antiferromagnetic (AF) long-range order.\cite{Hirsh85,White89} In the
weak-coupling limit ($U\ll 4t$), a Fermi surface instability gives rise to a
spin-density-wave ground state as first suggested by
Slater.\cite{Slater51}  The AF long-range
order produces a gap in the quasi-particle (QP) excitation spectrum so
that the system becomes insulating below the AF transition
temperature. In the strong-coupling regime ($U\gg 4t$), fermions are
localized by the strong Coulomb repulsion (Mott-Hubbard localization),
thus creating local (magnetic) moments on the lattice sites that
are well described by the Heisenberg model.\cite{Gebhard97,Auerbach94}
These local moments order at low temperature and give rise to a
Mott-Heisenberg antiferromagnet.  

The main difference between Slater and Mott-Heisenberg
antiferromagnets lies in the existence or absence of preformed local
(magnetic) moments above the N\'eel temperature
$T_{\rm N}$. \cite{Gebhard97} 
In the weak-coupling limit, we
expect a Fermi liquid phase down to temperatures very close to $T_{\rm N}$
where critical AF fluctuations start to grow. In the strong-coupling
limit, the system is insulating both above (Mott-Hubbard insulator)
and below (Mott-Heisenberg antiferromagnet) the N\'eel temperature. 

This simple view, while correct in 3D, breaks down
in 2D. In 2D systems, thermal (classical) fluctuations preclude 
a finite-temperature AF phase transition, and the phase
transition occurs at $T_{\rm N}=0$ 
in agreement with the Mermin-Wagner theorem.\cite{Mermin66}
Nevertheless, below a crossover temperature $T_X$, the system enters a
renormalized classical regime where 
AF fluctuations start to grow exponentially. Below $T_X$, the
Fermi-liquid description breaks down even at weak coupling, although
the system remains metallic. Instead of well-defined Landau's
QP's, the fermion spectral function ${\cal A}({\bf
k},\omega)$ exhibits two (broadened) peaks separated by a
pseudogap. 

The existence of a pseudogap at weak coupling is best understood by
considering the zero-temperature limit. At zero temperature, ${\cal A}({\bf
k},\omega)$ is expected to exhibit two peaks corresponding to the Bogoliubov
QP's as in the Hartree-Fock (HF) theory. These two peaks are separated by
the AF gap which is due to the presence of magnetic long-range
order. At any finite temperature, 
the AF long-range order disappears in 2D. However, by continuity, the
two-peak structure in ${\cal A}(\mathbf{k},\omega)$ cannot disappear as soon 
as we raise the temperature. As pointed out in
Ref.~\onlinecite{Vilk97}, the only possible scenario is that at finite
by low temperature, the fermion spectral function exhibits two
broadened peaks, which are precursors of the zero-temperature Bogoliubov QP's,
separated by a pseudogap. At strong coupling, the zero-temperature gap
survives at finite temperature since the system is a Mott-Hubbard
insulator.  

The simplest description of the AF ground state of the 2D half-filled
Hubbard model is based on the HF theory. It is known that
the HF theory remains meaningful even at large $U$. In
particular, spin-wave modes obtained from the Heisenberg model with
an exchange coupling $J=4t^2/U$ can be reproduced from a
random-phase-approximation (RPA) calculation about the AF HF
solution. \cite{Moriya85,Schrieffer89,Singh90,Chubukov92}   
The influence of the spin-wave modes on the fermionic excitations has
been studied within one-loop\cite{Vignale90} and self-consistent
one-loop\cite{Brenig93,Altmann95} approximations. A QP
picture for the coherent motion of a particle or a hole appears to be
still valid. However, AF quantum fluctuations lead to a significant
reduction of the Bogoliubov QP spectral weight, with a concomitant
redistribution of spectral intensity into incoherent
excitations, and a strong renormalization of the AF gap. These conclusions
are supported by numerical work on the Hubbard
model\cite{Dagotto92,Leung92,Feng92} and, in the strong-coupling
limit, by analytical or numerical analysis of the $t$-$J$
model. \cite{Dagotto94}    

In spite of its success at zero temperature, the HF theory
fails in 2D since it predicts AF long-range order at finite
temperature. In the weak-coupling limit, alternative approaches, which
do satisfy the Mermin-Wagner theorem, have been proposed: 
Moriya's self-consistent-renormalized theory,\cite{Moriya85,Moriya00,Moriya90}
the fluctuation exchange approximation (FLEX),
\cite{Bickers89}  or the two-particle self-consistent
theory. \cite{Vilk97}  None of these approaches gives a unified
description of the 
magnetic properties of the 2D Hubbard model at finite temperature,
both at weak and strong coupling. At strong coupling, in the
Mott-Hubbard insulating state, spin degrees of freedom are usually described
by the Heisenberg model for which various methods are
available. \cite{Chakravarty89,Chubukov94,Auerbach94}  

Beside their limitation to the weak-coupling regime, these approaches are
also unable to account for the strong suppression of the amplitude
fluctuations of the AF order parameter at low temperature and
therefore essentially describe {\it Gaussian} spin fluctuations. Below the
crossover temperature $T_X$, amplitude fluctuations are indeed frozen
and only directional fluctuations [i.e. (transverse) spin waves]
survive at low energy. The calculation of the single-particle Green's
function usually relies on a paramagnon-like self-energy describing
free fermions that couple to Gaussian order parameter
fluctuations. \cite{Bickers89,Langer95,Deisz96,Vilk97,Vilk96} This
kind  of approach has been originally introduced   
by Lee, Rice and Anderson to
explain the suppression of the density of states associated with order
parameter fluctuations near a charge-density-wave
instability. \cite{Lee73} It has been since studied by many authors,
in one and two dimensions. 
\cite{Sadovskii74,Sadovskii79,Tchernyshyov99,Bartosch99,Millis00,Schmalian98,Schmalian99,Kampf89,Kampf90,Monthoux93,Altshuler95,Vilk97a,Abanov00,Posazhennikova02}
The assumption of Gaussian spin 
fluctuations leads to an overestimation of the fermion density of
states at low energy. \cite{Monien01} 
Moreover, the artificial presence of
amplitude fluctuations does not allow to reach the correct $T\to 0$
limit. \cite{Tchernyshyov99,Dupuis02}
The effect of ``directional'' (i.e. phase) fluctuations of a
complex order parameter on the fermion density of states has been
studied both for incommensurate 1D Peierls systems
\cite{Monien01a,Bartosch00} and 2D superconductors.\cite{Loktev01} 
Transverse spin-wave fluctuations in the finite-temperature 
2D Hubbard model have not received as much attention so far.

On the experimental side, antiferromagnetism and pseudogaps are
ubiquitous in low-dimensional strongly correlated fermion
systems. Pseudogaps were first observed in quasi-1D systems near a
charge-density-wave instability. \cite{Lee73,Gruner} More recently, a
pseudogap has been observed in the metallic phase of high-$T_c$
superconductors.\cite{Timusk99,Tallon01} Whether the pseudogap in
these systems is of magnetic or pairing origin is still a matter of
intense debate.  

In this paper, we describe a theoretical approach which provides a
unified view of the 2D half-filled Hubbard model at low temperature
(including $T=0$) and for any value of the Coulomb
repulsion. \cite{KBND03} It is based on a non-linear sigma
model (NL$\sigma$M) description of spin fluctuations. At zero
temperature, our theory describes the evolution from a Slater ($U\ll 4t$)
to a Mott-Heisenberg ($U\gg 4t$) antiferromagnet. At finite
temperature, it predicts a pseudogap at weak-coupling due to strong AF
fluctuations, and a Mott-Hubbard gap at strong coupling. Since it
takes into account only directional fluctuations of the AF order
parameter, it is valid for $T\ll T_X$, where $T_X$ is a crossover
temperature which marks the onset of AF short-range order. In
Ref.~\onlinecite{Dupuis02}, one of the present authors reported a
calculation of the fermion spectral function in the weak-coupling
limit of the Hubbard model using a NL$\sigma$M description of spin
fluctuations. However, the limitations encountered by previous
approaches could not be overcome. 

As first shown by Schulz, \cite{Schulz95} spin fluctuations in the
2D Hubbard model at low
temperature can be described by a NL$\sigma$M for any value of the
Coulomb repulsion. \cite{Sengupta00} In 
Sec.~\ref{sec0derivation}, we
give a detailed derivation of the NL$\sigma$M starting from the
Hubbard model.  The parameters of the NL$\sigma$M, the bare spin stiffness
$\rho_s^0$ and the spin-wave velocity $c$, are calculated as a
function of the ratio $U/t$.  For $U\gg 4t$, we recover the
NL$\sigma$M derived from the Heisenberg model with an exchange
coupling $J=4t^2/U$. In
Sec.~\ref{sec0modele0sigma}, we introduce the CP$^1$ representation of the
NL$\sigma$M where the N\'eel field (giving the direction of the local
AF order) is expressed in terms of two
Schwinger bosons. This allows a simple saddle-point
solution\cite{Auerbach94} from which we obtain the magnetic phase
diagram of the 2D Hubbard model. At zero temperature, there is
condensation of the Schwinger bosons for any value of $U$, which
signals the presence of AF long-range order. At finite temperature,
the system is disordered by thermal fluctuations, but the AF
correlation length remains exponentially large below a crossover
temperature $T_X$ (renormalized classical
regime\cite{Chakravarty89}). In Sec.~\ref{sec0fc0spec}, we 
study the fermion spectral properties. The fermion is written as the
product of a Schwinger boson and a pseudo-fermion whose spin is
quantized along the (fluctuating) N\'eel field. Such a decomposition
is reminiscent of slave-boson\cite{Kotliar93} or slave-fermion
\cite{Jayaprakash89,Yoshioka89, Auerbach91} theories.\cite{note4} It
allows us to approximate the fermion Green's function by the product
(in direct space) of the Schwinger boson propagator 
(which is obtained from the NL$\sigma$M) and the HF fermionic
propagator. At weak-coupling ($U\ll 4t$) and zero temperature, our
results clearly describe a Slater antiferromagnet. The AF gap
$2\Delta_0 \sim te^{-2\pi\sqrt{t/U}}$ is exponentially small. As in the
HF theory, there are well-defined Bogoliubov QP's. However,
because of AF quantum fluctuations,
their spectral weight is reduced by a factor $n_0$ ($0<n_0<1$) which
is given by the fraction of condensed Schwinger bosons in the
ground state. The missing weight ($1-n_0$) is transferred to incoherent
excitations at higher energy ($1-n_0\ll 1$ when $U\ll 4t$). As $U$
increases, the AF gap increases 
and spectral weight is progressively transferred from the Bogoliubov
QP's to the incoherent excitation background. At strong coupling
($U\gg 4t$), our results are typical of a Mott-Heisenberg
antiferromagnet. The AF gap 
$2\Delta_0$ is of order $U$. The incoherent excitation background
carries a significant fraction of spectral weight (i.e. $n_0$
and $1-n_0$ are of the same order) and extends over an energy scale of
order $J=4t^2/U$ above the Bogoliubov QP energy $\pm E_{\bf k}$. At 
finite temperature, the Bogoliubov QP's disappear ($n_0=0$ in the
absence of Bose condensation) and only incoherent excitations
survive. Nevertheless, precursors of the 
zero-temperature Bogoliubov QP's show up as sharp peaks at $\pm
E_{\bf k}$ in the spectral function ${\cal A}(\mathbf{k},\omega)$, with a width
of order $T$. We show that these peaks continuously evolve into the
zero-temperature Bogoliubov QP peaks as $T\to 0$. This ensures that
the spectral function ${\cal A}(\mathbf{k},\omega)$ is continuous at
the $T_{\rm N}=0$ phase transition. The high-energy incoherent excitation
background is little affected by a 
finite temperature, but the presence of thermal AF fluctuations gives
rise to fermionic states below the zero-temperature AF gap
$\Delta_0$. At weak coupling, the gap is completely filled and
replaced by a pseudogap. At strong coupling, the zero-temperature gap
survives at finite temperature and the system is a Mott-Hubbard
insulator. 

On the basis of a numerical calculation in the framework of the
dynamical cluster approximation, Moukouri and Jarrell have called into
question the existence of a Slater scenario in the 2D half-filled
Hubbard model. \cite{Moukouri01,Kyung03,Anderson97} They argue that
the system is always 
a Mott-Hubbard insulator at low (but finite) temperature even at weak
coupling. We will show that their results are not in contradiction
with a Slater scenario at weak coupling, but merely reflect the strong
suppression of the density of states due to the pseudogap
(Sec.~\ref{subsec:mit}). 

At half-filling, the repulsive Hubbard model can be mapped exactly
onto the attractive model by a canonical transformation. 
\cite{Micnas90} This transformation maps the ${\bf
  q}=(\pi,\pi)$ spin correlations of the repulsive model
onto the $\mathbf{q}=0$ pairing and $\mathbf{q}=(\pi,\pi)$
charge correlations of the 
attractive model, but leaves the single-particle Green's function and
the spectral function ${\cal A}(\mathbf{k},\omega)$ invariant. Thus the results
obtained in this paper apply also to the attractive Hubbard model, but
with a different physical meaning (Sec.~\ref{sec0hubbard0attractif}). 
At zero temperature, there is superconducting and charge-density-wave
long-range orders. As the attractive interaction strength increases,
there is a smooth crossover from a BCS to a Bose-Einstein behavior. 
At finite temperature, the weak-coupling pseudogap is due to strong
pairing and charge fluctuations, whereas the strong-coupling gap is a
consequence of the presence of preformed particle-particle pairs.

\section{Derivation of the NL$\sigma$M} 
\label{sec0derivation}

The Hubbard model is defined by the Hamiltonian
\begin{equation}
		H 
	= 
		- \sum_{\mathbf{r} , \sigma} 
		c_{\mathbf{r} \sigma}^{\dagger} 
		( \hat{t} + \mu ) c_{\mathbf{r} \sigma}
		+ U \sum_{\mathbf{r}} 
		c_{\mathbf{r} \uparrow}^{\dagger}  c_{\mathbf{r} \uparrow}
		c_{\mathbf{r} \downarrow}^{\dagger} c_{\mathbf{r} \downarrow}
	,
	\label{eq0hamiltonian}
\end{equation}
where $\hat{t}$ is the nearest-neighbor hopping operator:
\begin{equation}
		\hat{t} c_{\mathbf{r}\sigma}
	= 
		t
		\left(
			c_{ \mathbf{r} + \mathbf{e}_x\sigma } + 
			c_{ \mathbf{r} - \mathbf{e}_x\sigma } +
			c_{ \mathbf{r} + \mathbf{e}_y \sigma} + 
			c_{ \mathbf{r} - \mathbf{e}_y \sigma}
		\right)
	\label{eq0t} .
\end{equation}
At half-filling the chemical potential $\mu$ equals $U/2$.
$e_x$ and $e_y$ denote unit vectors along the $x$ and $y$ directions.
$c_{\mathbf{r} \sigma}^{\dagger}$ ($c_{\mathbf{r} \sigma}$) creates (annihilates)
a fermion of spin $\sigma$ at the lattice site $\mathbf{r}$. We take the lattice
spacing equal to unity and set $\hslash = k_{\mathrm{B}} =1$ throughout the paper.

We can represent the partition function of the system as a path
integral over Grassmann fields
$\psi_{\mathbf{r} \sigma}$, $\psi_{\mathbf{r} \sigma}^{\star}$.
The action can be written as $S_{ \mathrm{kin} }+S_{ \mathrm{int} }$ with
\begin{eqnarray}
		S_{\mathrm{kin}} 
	&=& 
		\int_0^{\beta} d \tau \sum_{\mathbf{r}}
		\Psi_{\mathbf{r}}^{\dagger} 
		( \partial_{\tau} - \mu - \hat{t} ) 
		\Psi_{\mathbf{r}}
	\label{eq0act0kin} , \\
		S_{\mathrm{int}} 
	&=& U
		\int_0^{\beta} d \tau \sum_{\mathbf{r}}
		\psi_{\mathbf{r} \uparrow}^{\star}
		\psi_{\mathbf{r} \uparrow}
		\psi_{\mathbf{r} \downarrow}^{\star}
		\psi_{\mathbf{r} \downarrow}
	\label{eq0act0int} ,
\end{eqnarray} 
where $\beta=1/T$ is the inverse temperature.
In the kinetic action $S_{\rm kin}$ we have used the spinor representation
$\Psi=\left( \psi_{\uparrow} , \\ \psi_{\downarrow} \right)^T$.
To describe collective spin and charge fluctuations, we introduce auxiliary fields.
The standard approach is to write the interaction part of the action as
$
		\psi_{\mathbf{r} \uparrow}^{\star} 
		\psi_{\mathbf{r} \uparrow} 
		\psi_{\mathbf{r} \downarrow}^{\star}
		\psi_{\mathbf{r} \downarrow}
	= 
		\frac{1}{4} 
		\left( 
			\Psi_{\mathbf{r}}^{\dagger} 
			\Psi_{\mathbf{r}}  
		\right)^2 
	- 
		\frac{1}{4} 	
		\left( 
			\Psi_{\mathbf{r}}^{\dagger} 
			\sigma_3
			\Psi_{\mathbf{r}} 
		\right)^2 
$,
and to perform a Hubbard-Stratonovich transformation by means of two real auxiliary
fields $\Delta_{c \mathbf{r}}$ and $\Delta_{s \mathbf{r}}$. Although this procedure
recovers the standard mean-field (or HF) theory of the N\'eel state within
a saddle-point approximation, it leads to a loss of spin rotation invariance and
does not allow to obtain the spin-wave Goldstone modes. Fluctuations of 
$\Delta_{c \mathbf{r}}$ and $\Delta_{s \mathbf{r}}$ correspond to gapped amplitude
modes. Alternatively, one could write $S_{\mathrm{int}}$ in an explicitly 
spin-rotation invariant form, e.g. 
$
		\psi_{\mathbf{r} \uparrow}^{\star}
		\psi_{\mathbf{r} \uparrow} 
		\psi_{\mathbf{r} \downarrow}^{\star}
		\psi_{\mathbf{r} \downarrow}
	= 
		- 
		\frac{1}{6} 	
		\left( 
			\Psi_{\mathbf{r}}^{\dagger} 
			\boldsymbol{\sigma}
			\Psi_{\mathbf{r}} 
		\right)^2
$
($\boldsymbol{\sigma}=(\sigma_1,\sigma_2,\sigma_3)$ denotes the Pauli
		matrices), and use a vector Hubbard-Stratonovich 
field. Such decompositions, however, do not reproduce the HF results at the
saddle-point level.\cite{Schulz95} As noted earlier,\cite{Schulz95,Weng91} this difficulty can
be circumvented by using the decomposition
\begin{equation}
		\psi_{\mathbf{r} \uparrow}^{\star} 
		\psi_{\mathbf{r} \uparrow}
		\psi_{\mathbf{r} \downarrow}^{\star}
		\psi_{\mathbf{r} \downarrow}
	= 
		\frac{1}{4} 
		\left( 
			\Psi_{\mathbf{r}}^{\dagger} 
			\Psi_{\mathbf{r}}  
		\right)^2 
	- 
		\frac{1}{4} 	
		\left( 
			\Psi_{\mathbf{r}}^{\dagger} 
			\boldsymbol{\sigma} 
			\cdot
			\mathbf{\Omega}_{\mathbf{r}}
			\Psi_{\mathbf{r}} 
		\right)^2 ,
	\label{eq0psi}
\end{equation} 
where $\mathbf{\Omega}_{\mathbf{r}}$ is a site- and time-dependent unitary
vector. Spin-rotation invariance is made explicit by 
performing an angular integration over $\mathbf{\Omega}_\mathbf{r}$ 
at each site and time (with a measure normalized to
unity). The Hubbard-Stratonovich transformation then reads
\begin{eqnarray}
	e^{ - S_{\mathrm{int}} } 
	&=& 
		\int \mathcal{D} [ \Delta_c , \Delta_s , \mathbf{\Omega} ]
	\label{eq0hs} \nonumber \\
	&&
		\times
		e^{ 
			- \int_0^{\beta} d \tau \sum_{\mathbf{r}} 
			\left[	
				\frac{1}{U} 
				( 	
					\Delta_{c \mathbf{r}}^2 
					+ \Delta_{s \mathbf{r}}^2 
				)
				-
				\Psi_{\mathbf{r}}^{\dagger}
				(
					i \Delta_{c \mathbf{r}} 
					+ \Delta_{s \mathbf{r}} 
					\boldsymbol{\sigma} \cdot
		\mathbf{\Omega}_{\mathbf{r}} 
				)
				\Psi_{\mathbf{r}}
			\right]
		} . 
\end{eqnarray}	
Eq.~(\ref{eq0hs}) corresponds to an "amplitude-direction" representation, where
the magnetic order parameter field is given by 
$\Delta_{s \mathbf{r}} \mathbf{\Omega}_\mathbf{r}$.
The HF theory is now recovered from a saddle-point approximation over the
auxiliary fields $\Delta_{c \mathbf{r}}$, $\Delta_{s \mathbf{r}}$ and 
$\mathbf{\Omega}_{\mathbf{r}}$ (Sec.~\ref{sec0hartree0fock}). Spin-wave
excitations can then be obtained by considering small fluctuations of the 
$\mathbf{\Omega}_{\mathbf{r}}$ field about its saddle-point value. In 
Sec.~\ref{sec0modele0sigma} we show that the amplitude-direction representation
(\ref{eq0hs}) allows to go beyond the N\'eel-ordered HF state and derive an 
effective action for the $\mathbf{\Omega}_{\mathbf{r}}$ field.

\subsection{HF theory} \label{sec0hartree0fock}

Making the ansatz of an antiferromagnetic order, i.e. taking constant auxiliary fields 
$ \Delta_{c \mathbf{r}} = \Delta_c$, $\Delta_{s \mathbf{r}} = \Delta$ and
a staggering vector 
$\mathbf{\Omega}_\mathbf{r} = (-1)^\mathbf{r} \mathbf{u}_z$
parallel to the $z$ axis,
one obtains the saddle-point equations
\begin{eqnarray}
		i \Delta_{c} 
	&=&
		-\frac{U}{2}
		\left<
			\Psi_{\mathbf{r}}^{\dagger}
			\Psi_{\mathbf{r}}
		\right>
	\label{eq0selle0charge} , \\
		\Delta
	&=&
		\frac{U}{2}
		(-1)^\mathbf{r}
		\left<
			\Psi_{\mathbf{r}}^{\dagger}
			\sigma_3
			\Psi_{\mathbf{r}}
		\right>
	\label{eq0selle0spin} .
\end{eqnarray}		
At half-filling, the saddle-point value $i\Delta_c =-U/2$ cancels the chemical 
potential term in (\ref{eq0act0kin}). The HF action is quadratic,
\begin{equation}
		S_{ \mathrm{HF} } 
	=
		\int_0^{\beta} d \tau
		\sum_{\mathbf{r}}
		\Psi_{\mathbf{r}}^{\dagger} 
		\left(
			\partial_{\tau} - \hat{t}  
			- (-1)^{\mathbf{r}} \Delta \sigma_3
		\right)
		\Psi_{\mathbf{r} }
	\label{eq0act0hf} ,
\end{equation}	
and can easily be diagonalized. Due to the translational symmetry breaking
there is unit cell doubling. In the reduced Brillouin zone scheme
($|k_x| + |k_y| \leq \pi$) elementary excitations are exhausted by two
bands of Bogoliubov QP's at energies 
$\pm E_{\mathbf{k}} = \pm \sqrt{\Delta^2 + \epsilon_{\mathbf{k}}^2}$,
$\epsilon_{\mathbf{k}} = -2t (\cos k_x + \cos k_y)$ being the energy
of free fermions. 

Using the HF action (\ref{eq0act0hf}) one obtains the HF 
single-particle Green's function
\begin{eqnarray}
		- \left< 
			\phi_{\mathbf{k} \omega \sigma}
			\phi_{\mathbf{k'} \omega' \sigma'}^{\star}  
		\right> 
	&=&
		\delta_{ \omega , \omega' }
		\delta_{ \sigma , \sigma' }
		\mathcal{G}_{\sigma}^{\mathrm{HF}}( \mathbf{k} , \mathbf{k}' , \omega )
	\label{eq0prop0hf} , \\ 
		\mathcal{G}_{\sigma}^{\mathrm{HF}}(\mathbf{k} , \mathbf{k}' , \omega )
	&=&
		-
		\delta_{ \mathbf{k} , \mathbf{k}' } 
		\frac	{ i \omega + \epsilon_\mathbf{k} }
			{ \omega^2 + E_\mathbf{k}^2 }
		+
		\delta_{ \mathbf{k} , \mathbf{k}'+ \boldsymbol{\pi} } 
		\frac	{ \sigma  \Delta }
			{ \omega^2 + E_\mathbf{k}^2 } ,
	\nonumber \\
	\label{eq0green0hf} .
\end{eqnarray}
where $\boldsymbol{\pi}=(\pi,\pi)$ and $\omega\equiv (2n+1)\pi T$ ($n$ integer)
is a fermionic Matsubara frequency.
The propagator (\ref{eq0prop0hf}) makes it in turn possible to give an explicit
form to the gap equation (\ref{eq0selle0spin}):
\begin{eqnarray}
		\frac{1}{U} 
	= 
		\int_\mathbf{k} 
		\frac	{ \tanh( \frac{\beta E_\mathbf{k}}{2} ) }
			{2 E_\mathbf{k}} .
	\label{eq0gap}
\end{eqnarray}	 
We use the notation
$\int_\mathbf{k} = \int_{-\pi}^{\pi} \int_{-\pi}^{\pi} 
\frac{d k_x}{2 \pi} \frac{d k_y}{2 \pi}$.
Eq.~(\ref{eq0gap}) predicts a phase transition at a finite temperature 
$T_{\mathrm{N}}^{\mathrm{HF}}$, which is exponentially small at weak coupling
and approaches $U/4$ at strong coupling. Similarly to the transition
temperature, the zero-temperature gap
$\Delta_0$ tends to $U/2$ at strong coupling and is exponentially
small at weak coupling: 
\begin{equation}
	\Delta_0 \simeq 32 t e^{ -2 \pi \sqrt{t/U}}
	\label{eq0delta0faible} .
\end{equation}

\subsection{Spin fluctuations}
\label{sec0fluctuations0spin}

In 2D, the HF theory breaks down at finite temperature, since it predicts
AF long-range order below $T_{\mathrm{N}}^{\mathrm{HF}}$. Nevertheless, the
HF transition temperature bears a physical meaning as a crossover
temperature below which the amplitude of the AF order parameter takes
a well-defined 
value. This is sometimes interpreted as the appearance of local moments with
an amplitude $\Delta_0/U$. Note that at weak coupling the "local" moments
can be defined only at length-scale of order $\xi_0\sim t/\Delta_0$,
which corresponds to the size of bound particle-hole pairs in the HF 
ground state. Thus, {\it stricto sensu}, local moments form only in
the strong-coupling limit when $\xi_0\sim 1$.

Below $T_{\mathrm{N}}^{\mathrm{HF}}$, it is natural to neglect
amplitude fluctuations of the AF order parameter and derive an
effective action of the $\mathbf{\Omega}_\mathbf{r}$ field by
integrating out the fermionic degrees 
of freedom. We call $T_{\mathrm{X}}$ the crossover temperature below
which AF short-range order appears. As will be shown subsequently, in the weak
coupling limit $T_{\mathrm{X}} \sim T_{\mathrm{N}}^{\mathrm{HF}}$, whereas at
strong coupling $T_{\mathrm{X}}\sim J=4t^2/U \ll
T_{\mathrm{N}}^{\mathrm{HF}}$. For $T\ll T_X$, the amplitude of the AF
order parameter can be 
approximated by its zero-temperature HF value $\Delta_0$. Following 
Haldane,\cite{Haldane83,Auerbach94} in the presence of AF short-range order 
($T \lesssim T_{\mathrm{X}}$)
we write
\begin{equation}
	\mathbf{\Omega}_{\mathbf{r} } =
		(-1)^{\mathbf{r}}
		\mathbf{n}_{\mathbf{r} } 
		\sqrt{ 1 - \mathbf{L}_{\mathbf{r} }^2 }
	+  \mathbf{L}_{\mathbf{r} }.
	\label{eq0haldane}
\end{equation} 
$\mathbf{n}$ is a unitary vector representing the N\'eel field, whereas
$\mathbf{L}$ is the canting vector, orthogonal to $\mathbf{n}$, taking account of local 
ferromagnetic fluctuations.  $\mathbf{n}$ is assumed to be slowly-varying and $\mathbf{L}$
to be small. We perform at each site and time a rotation in spin
space and introduce a new fermionic field 
$\Phi_\mathbf{r}$ defined by $\Psi_\mathbf{r} = 
R_\mathbf{r} \Phi_\mathbf{r}$. $R_\mathbf{r}$ is a time- and
site-dependent $\mathrm{SU}(2)/\mathrm{U}(1)$ matrix satisfying  
\begin{equation}
	\mbox{\boldmath$\sigma$} \cdot \mathbf{n}_\mathbf{r}
	=
	R_{ \mathbf{r} } \sigma_3 
	R_{ \mathbf{r} }^{\dagger} . 
	\label{eq0rotation}
\end{equation}
The above definition means that $\mathcal{R}_\mathbf{r}$, the SO(3) element associated 
to $R_\mathbf{r}$, maps $\mathbf{u}_z$ onto $\mathbf{n}_\mathbf{r}$.
The $\mathrm{U}(1)$ gauge freedom is due to rotations around the $z$ axis, which do not
change the physical state of the system. The spin of the 
pseudo-fermions $\Phi_\mathbf{r}$ is quantized along the $\mathbf{n}_\mathbf{r}$ 
axis. In order to express the action in terms of the new spinor variable,
it is convenient to make use of the $\mathrm{SU}(2)$ gauge field 
$A_{\mu \mathbf{r}} = \sum_{\nu = 1,2,3} A_{\mu \mathbf{r}}^{\nu} \sigma_{\nu}$
defined as
\begin{eqnarray}
		A_{0 \mathbf{r}}
	&=&
		- R_{\mathbf{r}}^{\dagger} 
		\partial_{\tau} 
		R_{\mathbf{r}}
	\label{eq0jauge0temps} , \\
		A_{\mu \mathbf{r}}
	&=& 
		i  R_{\mathbf{r}}^{\dagger} 
		\partial_{\mu} 
		R_{\mathbf{r}}
		, \qquad \mu = x , y
	\label{eq0jauge0espace} .
\end{eqnarray} 
We also define the rotated canting field
$\mathbf{l}_\mathbf{r} = \mathcal{R}_\mathbf{r}^{-1} \mathbf{L}_\mathbf{r}$.
Given that 
$\mathcal{R}_\mathbf{r}^{-1} \mathbf{n}_\mathbf{r} = \mathbf{u}_z$
and $\mathbf{L}_\mathbf{r} \perp \mathbf{n}_\mathbf{r}$,
the $\mathbf{l}_\mathbf{r}$ vector lies in the $x-y$ plane.
Using the identity
\begin{eqnarray}
		\Phi_{\mathbf{r}}^{\dagger} 
		R_{\mathbf{r}}^{\dagger} 
		R_{ \mathbf{r} + \mathbf{e}_{\mu} }
		\Phi_{\mathbf{r} + \mathbf{e}_{\mu} }
	=	
		\Phi_{\mathbf{r}}^{\dagger} 
		e^{ \partial_\mu - i A_{ \mu \mathbf{r} } }
		\Phi_{\mathbf{r}}
	\label{eq0der0spatiale}
\end{eqnarray}	 
we re-express the kinetic and interaction parts of the action as
\begin{eqnarray}
		S_{\mathrm{kin}}
	&=&
		\int_0^{\beta}d\tau \sum_{\mathbf{r}}
		\Phi_{\mathbf{r} }^{\dagger} 
		\big[
			\partial_{\tau} -  A_{0 \mathbf{r}}
	\nonumber \\
	&&		-2 t \sum_{\mu = x,y}
			\cos( -i \partial_{\mu} - A_{\mu \mathbf{r}} ) 
		\big]
		\Phi_{\mathbf{r} } 
	\label{eq0act0cin0phi} , \\
		S_{\mathrm{int}} 
	&=&	- \Delta_0
		\int_0^{\beta} d \tau \sum_{\mathbf{r}}
			\Phi_{ \mathbf{r} }^{\dagger}
			[
				(-1)^{\mathbf{r}} 
				\sigma_3
				\sqrt{ 1 - \mathbf{l}_{\mathbf{r} }^2 }
		                    \nonumber \\  &&	+ 
				\mathbf{l}_{\mathbf{r} } 
				\cdot
				\mbox{\boldmath$\sigma$} ]  
			\Phi_{ \mathbf{r} }
	\label{eq0act0int0phi} .
\end{eqnarray}

In the above expressions, both $\mathbf{l}$ and $A_\mu$ are small,
since the gauge field is  
of the order of $\partial_\mu \mathbf{n}$. We expand 
Eqs.~(\ref{eq0act0cin0phi}-\ref{eq0act0int0phi})
to second order in these variables. To zeroth order, we recover the HF action 
$S_{\mathrm{HF}}[\Phi]$ defined in (\ref{eq0act0hf}). The first- and second-order 
corrections in $A_{\mu}^{\nu}$ yield paramagnetic and diamagnetic terms 
$S_{\mathrm{p}}$ and $S_{\mathrm{d}}$, respectively. The corrections
in $\mathbf{l}$ give first- and  
second-order ferromagnetic fluctuations  $S_{l}$ and $S_{l^2}$:\cite{note6}
\begin{eqnarray}
		S_{ \mathrm{p} } 
	&=&
		- \int_0^{\beta} d \tau 
		\sum_{\mu=0,x,y \atop {\nu=1,2,3 \atop \mathbf{r}} }
		j_{\mu \mathbf{r}}^{\nu} A_{\mu \mathbf{r}}^{\nu}
	\label{eq0act0para} , \\
		S_{ \mathrm{d} } 
	&=&
		\frac{t}{2} \int_0^{\beta} d \tau 
		\sum_{\mu=x,y \atop {\nu=1,2,3 \atop \mathbf{r}}} 
                {A_{\mu \mathbf{r}}^\nu}^2
		\Phi^{\dagger}_{\mathbf{r}} 
			\cos(-i \partial_{\mu}) 
		\Phi_{\mathbf{r} }
		+ \mathrm{c.c.}
	\label{eq0act0dia} , \\
		S_{l} 
	&=&
		- \Delta_0 \int_0^{\beta} d \tau 
		\sum_{\nu=1,2 \atop \mathbf{r}}
		l_{\mathbf{r} }^{\nu} j_{0 \mathbf{r}}^{\nu}
	\label{eq0act0l} , \\
		S_{l^2}
	&=&
		 \frac{\Delta_0}{2} \int_0^{\beta} d \tau \sum_{\mathbf{r}}
		(-1)^{\mathbf{r}} \mathbf{l}_{\mathbf{r} }^2  j_{0 \mathbf{r}}^{3}
	. \label{eq0act0l02}
\end{eqnarray} 
The spin-density currents $j_{\mu}^{\nu}$ are defined by
\begin{eqnarray}
		j_ { 0 \mathbf{r} }^{\nu} 
	&=&
		\Phi_{\mathbf{r}}^{\dagger}  
		\sigma_{\nu} 
		\Phi_{\mathbf{r}},
	\label{eq0j0temps} \\
		j_{\mu \mathbf{r} }^{\nu} 
	&=&
		t \Phi_{\mathbf{r}}^{\dagger} 
		\sin(-i \partial_{\mu} ) 
		\sigma_{\nu} 
		\Phi_{\mathbf{r}}
		+ \mathrm{c.c.}, \,\,\,\,\,\,\,\, \mu=x,y. 
	\label{eq0j0espace}
\end{eqnarray} 

We now derive an effective action for the spin variables $\mathbf{n}$ and
$\mathbf{L}$ by integrating out the fermions. Keeping terms up to
second order in $A_\mu^\nu$ and ${\bf l}$, the effective action is
given by first- and second-order cumulants of  
the four perturbative terms $S_{\mathrm{p}}$, $S_{\mathrm{d}}$,
$S_{l}$, $S_{l^2}$ with 
respect to the HF action:
\begin{eqnarray}
		S_{\mathrm{eff}}[ \mathbf{n} , \mathbf{L} ]
	&=&
		\left< S_{\mathrm{p}} \right> 
		+ \left< S_{\mathrm{d}} \right>
		+ \left< S_{l} \right> + \left< S_{l^2} \right>
	\nonumber \\
	&&	- \frac{1}{2} \left< S_{\mathrm{p}}^2 \right>_c
		- \frac{1}{2} \left< S_{l}^2 \right>_c
		- \left< S_{\mathrm{p}} S_{l} \right>_c .
	\label{eq0act0cumul}
\end{eqnarray} 
Evaluation of the first-order cumulants is straightforward. Defining 
$\epsilon_c$ as the absolute value of the (negative) kinetic energy 
per site in the HF ground state we have
\begin{eqnarray}
		\left< S_\mathrm{p} \right>
	&=&
		- \frac{2 \Delta_0}{U} 
		\int_0^{\beta} d \tau 
		\sum_{\mathbf{r}}
		(-1)^{\mathbf{r}} A_{0 \mathbf{r}}^3
	\label{eq0cumul0s0p} , \\
		\left< S_\mathrm{d} \right> 
	&=&
		\frac{\epsilon_c}{4}  
		\int_0^{\beta} d \tau \sum_{\mu=x,y \atop {\nu=1,2,3 \atop \mathbf{r}} }
		{ A_{\mu \mathbf{r}}^{\nu} }^2
	\label{eq0cumcul0s0d} , \\
		\left< S_l \right> 
	&=&
		0
	\label{eq0cumul0s0l} , \\
		\left< S_{l^2} \right>
	&=&
		\frac{\Delta_0^2}{U} 
		\int_0^{\beta} d \tau 
		\sum_{\mathbf{r}}
		\mathbf{l}_{ \mathbf{r} }^2
	\label{eq0cumul0s0l02} .
\end{eqnarray} 
We recognize in equation (\ref{eq0cumul0s0p}) the usual Berry phase term. 
Since it is believed to play no role in a two-dimensional
antiferromagnet \cite{Auerbach94} we will ignore it in the following. 

The calculation of the second-order cumulants seems cumbersome at
first sight, since it involves (after moving to the Fourier space) the
current-current correlation function  
$
\Pi_{\mu \mu'}^{\nu \nu'}
( \mathbf{q} , \omega_\nu ; \mathbf{q'} , \omega_\nu') =
\langle 
	j_{\mu}^{\nu}(\mathbf{q} , \omega_\nu ) 
	j_{\mu'}^{\nu'}(\mathbf{q'} , \omega_\nu') 
\rangle_\mathrm{HF}
$. [$\omega_\nu=\nu2\pi T$ ($\nu$ integer) is a bosonic Matsubara frequency.]
In fact, as the correlator stands in front of second-order quantities,
we are interested 
only in its zero-frequency, zero-momentum value
$\Pi_{\mu \mu'}^{\nu \nu'}( \mathbf{0} , 0 ; \mathbf{0} , 0 )$
which we denote by
$\Pi_{\mu \mu'}^{\nu \nu'}$.
With the exception of
$\Pi_{0 0}^{1 1} = \Pi_{0 0}^{2 2}$
and
$\Pi_{x x}^{3 3} = \Pi_{y y}^{3 3}$
all these quantities vanish (see Appendix \ref{app0correltateur}),
so that we obtain  
\begin{eqnarray}
		\left< S_{\mathrm{p}}^2 \right>_c 
	&=&
		\Pi_{0 0}^{1 1}
		\int_{0}^{\beta} d \tau
		\sum_{ \mathbf{r} \atop \nu=1,2 }
			{A_{0 \mathbf{r}}^{\nu}}^2
	\nonumber \\
	&&+ \;
		\Pi_{x x}^{3 3}
		\int_{0}^{\beta} d \tau
		\sum_{ \mathbf{r} \atop \mu=x,y }
			{A_{\mu \mathbf{r}}^{3}}^2
	\label{eq0cumul0sp0sp} , \\
		\left< S_{l}^2 \right>_c
	&=&
		\Delta_0^2 \Pi_{0 0}^{1 1}
		\int_{0}^{\beta} d \tau
		\sum_{ \mathbf{r} \atop \nu=1,2 }
			{l_\mathbf{r}^{\nu}}^2 
	\label{eq0cumul0sl0sl} , \\
		\left< S_{\mathrm{p}} S_{l} \right>_c
	&=&
		\Delta_0 \Pi_{0 0}^{1 1}
		\int_{0}^{\beta} d \tau
		\sum_{ \mathbf{r} \atop \nu=1,2 }
		A_{0 \mathbf{r}}^{\nu}
		l_\mathbf{r}^{\nu}
	\label{eq0cumul0sp0sl} .
\end{eqnarray} 
Using the invariance of the current-current correlation function with
respect to rotations of axis $\mathbf{u}_z$, one can establish the identity
$\epsilon_c/2 - \Pi_{xx}^{33} = 0$.
It implies that the ${A_{\mu}^{3}}^2$
terms in the first- and second-order cumulants cancel each other,
which ensures the $\mathrm{U}(1)$ gauge invariance. The only remaining
correlator is the transverse spin  
susceptibility $\Pi_{0 0}^{1 1} \equiv \chi^{\perp}$. In order to express the
effective action $S_{\mathrm{eff}}[ \mathbf{n} , \mathbf{L} ]$
[Eq.~(\ref{eq0act0cumul})] 
in terms of $\mathbf{n}$ and $\mathbf{L}$ we use the relations (see
Appendix \ref{app0jauge})
\begin{eqnarray}
		\sum_{\nu=1,2} 
		{ A_{\mu \mathbf{r}}^{\nu} }^2
	&=&
		\zeta \frac{1}{4} 
		( \partial_{\mu} \mathbf{n}_\mathbf{r} )^2
	\label{eq0a0et0n01} , \\
		\sum_{\nu=1,2} 
		A_{0 \mathbf{r}}^{\nu}  l_\mathbf{r}^{\nu}
	&=&
		 \frac{i}{2} 	
		( 
			\mathbf{n}_\mathbf{r} 
			\wedge \partial_{\tau} 
			\mathbf{n}_\mathbf{r} 
		)
		\cdot \mathbf{L}_\mathbf{r}
	\label{eq0a0et0n02} ,
\end{eqnarray}
with $\zeta=1$ for $\mu=x,y$ and $\zeta=-1$ for $\mu=0$.
Putting everything together, we obtain the effective action
\begin{eqnarray}
	S_{\mathrm{eff}} &=&
	\frac{1}{2} \int_0^{\beta} d \tau \sum_{\mathbf{r}}
	\Big[
		\frac{\chi^{\perp}}{4} \dot{\mathbf{n}}_{\bf r}^2
		+ 
		\frac{\epsilon_c}{8} \sum_{\mu = x,y}
		( \partial_{\mu} \mathbf{n}_{\bf r} )^2
	\nonumber \\	
	&&	+ \Delta_0^2 \Big( \frac{2}{U}-\chi^{\perp} \Big)
	           \mathbf{L}_{\bf r}^2 
		- i \Delta_0 \chi^{\perp} ( \mathbf{n}_{\bf r}
	\wedge \dot \mathbf{n}_{\bf r} ) 
		\cdot \mathbf{L}_{\bf r}
	\Big]
	\label{eq0act0l0n}
\end{eqnarray}
where $ \dot{\mathbf{n}} = \partial_\tau \mathbf{n}$. Integrating out
the canting field with the constraint ${\bf L_r} \perp {\bf n_r}$, we
eventually obtain a  NL$\sigma$M for the N\'eel field: 
\begin{equation}
		S_{ \mathrm{NL}\sigma\mathrm{M} }[\mathbf{n}] 
	=
		\frac{\rho_s^0}{2} 
		\int_0^{\beta} d \tau \int d^2 r 
		\left[
			\frac{1}{c^2}  \dot\mathbf{n}_{\bf r}^2
			+ 
			\sum_{\mu = x,y}
			( \partial_{\mu} \mathbf{n}_{\bf r} )^2
		\right]
	\label{eq0act0msnl} ,
\end{equation}
where we have taken the continuum limit in real space. 
The bare spin stiffness $\rho_s^0$ and the spin-wave velocity $c$ are given by 
\begin{equation}
	\rho_s^0 = \frac{\epsilon_c}{8} , \qquad 
	c^2 = \frac{\epsilon_c}{2} \left( \frac{1}{\chi^{\perp}} -
	\frac{U}{2} \right) . 
	\label{eq0rho0et0c}
\end{equation} 
Eq.~(\ref{eq0act0msnl}) must be supplemented with a cutoff $\Lambda$
in momentum 
space. In the strong-coupling limit, where AF fluctuations are due to
local moments, 
the cutoff $\Lambda$ can be taken of the order of unity. In the
weak-coupling limit, 
the N\'eel field is ill-defined at length-scales smaller than $\xi_0
\sim t/\Delta_0$, 
since below $\xi_0$ "local" moments cannot be defined (see the
discussion at the beginning of Sec.~\ref{sec0fluctuations0spin}). We
therefore choose $\Lambda \sim \mathrm{min}( 1 , 2\Delta_0/c
)$.\cite{note7}  

\begin{figure}
	\centerline{\psfig{file=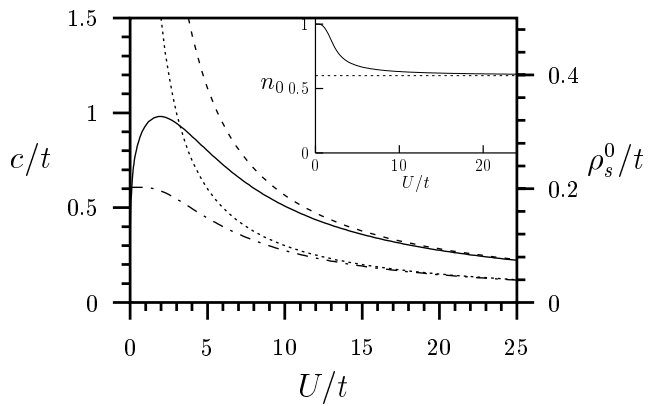,width=8.5cm,angle=0}}
	\caption{Spin-wave velocity $c$ (solid line) and bare spin stiffness 
	$\rho_s^0$ (dot-dashed line) {\it vs} $U$. For $U \gg 4t$ we recover 
	the results obtained from the Heisenberg model with $J= 4 t^2/U$
	(dashed line: spin-wave velocity, dotted line:
	bare spin stiffness). Inset: fraction of condensed bosons at $T=0$. 
	$1-n_0$ is exponentially small at weak coupling ($U \ll 4t$), while 
	$n_0 \simeq 0.6$ for $U \gg 4t$. $n_0$ determines the mean-value of the
	N\'eel field in the ground state  
	($\left<\mathbf{n}_\mathbf{r} \right> = n_0 \mathbf{u}_z$) 
	and the spectral weight of the Bogoliubov QP's 
	(see Sec.~\ref{sec0fc0spec}).}
\label{fig0coefs}
\end{figure}

For numerical computation of the spin-wave velocity and spin stiffness we use
the expressions
\begin{eqnarray}
	\chi^{\perp} &=& 2\Delta_0^2
	\int_{0}^{4 t} 	\frac	{ \rho_0(\epsilon) d \epsilon }
				{ ( \Delta_0^2 + \epsilon^2 )^{\frac{3}{2}} }
	, \\ 
	\rho_s^0 &=& 
	\frac{1}{4} \int_{0}^{4 t} 
		\frac 	{ \epsilon^2 \rho_0(\epsilon) d \epsilon }
			{ \sqrt{ \Delta_0^2 + \epsilon^2 } } .
	\label{eq0coefs}
\end{eqnarray}
$\rho_0$ is the density of states of free fermions on a square lattice. It can
be expressed, using the complete elliptic integral of the first kind $K$, as
$\rho_0( \epsilon )  = (2\pi^2 t)^{-1}K[(1-\epsilon^2/16 t^2)^{1/2}]$
for $|\epsilon|\leq 4t$.
In the strong coupling limit we recover the results obtained from the
Heisenberg model with an exchange coupling $J = 4t^2/U$: the spin
stiffness equals $J/4$ and the spin-wave velocity $\sqrt{2} J$. 
At weak coupling, $c$ goes to zero like 
$2\pi^{-1/2} t (U/t)^\frac{1}{4}$ and $\rho_s^{(0)}\sim t$. The factor
$(U/t)^\frac{1}{4}$ is due to the Van-Hove
singularities. The results are shown in Fig.~\ref{fig0coefs}.

The NL$\sigma$M defined by Eqs.~(\ref{eq0act0msnl}-\ref{eq0coefs}) was first
obtained by Schulz.\cite{Schulz95} The value of the spin-wave velocity
agrees with the result obtained from an RPA calculation about the
zero-temperature AF HF state.
\cite{Schrieffer89,Singh90,Chubukov92}

Note that we also expect a damping term with a characteristic
frequency $\omega_{\rm sf}$ in the  NL$\sigma$M action
(\ref{eq0act0msnl}) at weak coupling. This term comes from the damping
of spin fluctuations by fermion excitations which are gapless in the
weak coupling limit (see Sec.~\ref{sec0fc0spec}). It is  
missed in our approach since we expand around the zero-temperature AF
state which has only gapped quasi-particle excitations. In the
renormalized classical regime, this term is however negligible since
$\omega_{\rm sf}\propto \xi^{-2}\to 0$ (critical slowing down). 
\cite{Vilk97,Vilk97a}

\section{Magnetic phase diagram}
\label{sec0modele0sigma}

Let us recast the NL$\sigma$M action in a more usual form, by making use
of the coupling constant $g=c/\rho_s^0$:
\begin{equation}
		S_{ \mathrm{N L} \sigma \mathrm{M} }[\mathbf{n}]
	=
		\frac{1}{2 g} 
		\int_0^{\beta} d \tau \int d^2 r 
		\left[
				\frac{1}{c}  \dot\mathbf{n}_{\bf r}^2
			+ 
				c \sum_{\mu = x,y}
				( \partial_{\mu} \mathbf{n}_{\bf r} )^2
		\right] .
	\label{eq0act0msnl0g}
\end{equation}
To solve the NL$\sigma$M, we use
a saddle-point approximation in the $\mathrm{CP}^1$
representation,\cite{Auerbach94}  which proves 
well suited for the computation of the fermion Green's function.
In the $\mathrm{CP}^1$ representation, the N\'eel field is expressed in terms of
two Schwinger bosons,
\begin{equation}
	\mathbf{n}_{\mathbf{r}} 
	= 
		\mathrm{z}_{\mathbf{r}}^{\dagger} 
		\mbox{\boldmath $\sigma$} 
		\mathrm{z}_{\mathbf{r}}
	\label{eq0n0z} ,
\end{equation} 
with
$\mathrm{z}_{\mathbf{r}} = (z_{\mathbf{r} \uparrow} , z_{\mathbf{r}
  \downarrow})^T$. 
The condition $\mathbf{n}_\mathbf{r}^2 = 1$ translates into
$\mathrm{z}_{\mathbf{r}}^{\dagger} \mathrm{z}_{\mathbf{r}} = 1$. The rotation
matrix $R$ can be expressed as
\begin{equation}
	R_{\mathbf{r}} = 
		\left( 
		\begin{array}{cc}
			z_{\mathbf{r} \uparrow} & -z_{\mathbf{r}
			\downarrow}^{\star} \\ 
			z_{\mathbf{r} \downarrow}  & z_{\mathbf{r}
			\uparrow}^{\star}   
		\end{array}
		\right) .
	\label{eq0r0z}
\end{equation} 
The $\mathrm{U}(1)$ gauge symmetry now manifests itself in the
invariance of the  
$\mathbf{n}_\mathbf{r}$ vector and the relation (\ref{eq0rotation})
defining the rotation matrix under the transformation 
$\mathrm{z}_\mathbf{r} \rightarrow e^{ i \alpha_\mathbf{r} }
\mathrm{z}_\mathbf{r}$.  

The NL$\sigma$M expressed in terms of Schwinger bosons involves terms
quadratic and quartic in $\mathrm{z}_\mathbf{r}$. The latter turns out
to be proportional to ${A_{\mu \mathbf{r}}^z}^2$ with 
$A_{\mu \mathbf{r}}^z$ expressed in terms of Schwinger bosons
[Eqs.~(\ref{eq0jauge0espace})  
and (\ref{eq0r0z})]. It is decoupled by an auxiliary field $a_{\mu
\mathbf{r}}$. To handle the unimodularity condition
$\mathrm{z}_{\mathbf{r}}^{\dagger} \mathrm{z}_{\mathbf{r}} = 1$
one introduces Lagrange multipliers $\lambda_\mathbf{r}$ at 
each time and site. The partition function then becomes (see
Ref.~\onlinecite{Auerbach94}) 
\begin{eqnarray}
		Z
	&=&
		\int \mathcal{D}[ \mathrm{z} , a_\mu ,\lambda ]
		 e^{-S}
	 \\
		S
	&=&
	 \int_{0}^{\beta} d\tau \int d^2 r 
		\Big[
			i \lambda_\mathbf{r}
			(
				\mathrm{z}_\mathbf{r}^{\dagger} 
				\mathrm{z}_\mathbf{r}
				- 1
			)
			+
			\frac{2}{g c}
			\left|
				( \partial_\tau - a_{0 \mathrm{r}} )
				\mathrm{z}_\mathbf{r}
			\right|^2
	\nonumber \\
	&& + \frac{2 c}{g}
			\sum_{\mu = x,y}
			\left|
				( \partial_\mu - i a_{\mu \mathbf{r}} )
				\mathrm{z}_\mathbf{r}
			\right|^2
		\Big]
	\label{eq0act0z0lambda} ,
\end{eqnarray}	 
the $z_{\mathbf{r} \uparrow}$ and $z_{\mathbf{r} \downarrow}$ being now unconstrained 
bosonic fields. One then performs a saddle-point approximation over the 
$\lambda_\mathbf{r}$ and $a_{\mu \mathbf{r}}$  fields. When the
$\mathrm{CP}^1$ representation  
is generalized to the $\mathrm{CP}^{N-1}$ representation by introducing $N$ different $z$ bosons, 
the approximation becomes exact in the limit $N \rightarrow
\infty$. \cite{Auerbach94} Within the ansatz
of a uniform static saddle-point solution $i \lambda_\mathbf{r}=2 m^2/g c$ and
$a_{\mu {\bf r}} = 0$, the propagator can be read off from
(\ref{eq0act0z0lambda}):
\begin{eqnarray}
	&& -
		\left< 	
			z_{\mathbf{q} \, \omega_{\nu} \sigma} \,
			z_{ \mathbf{q'} \, \omega_{\nu}' \sigma'}^{\star} 
		\right>
	= 	
		\delta_{\mathbf{q} , \mathbf{q'}}
		\delta_{\omega_{\nu} , \omega_{\nu}'}  
		\delta_{\sigma , \sigma'}  
		\mathcal{D}_\sigma (\mathbf{q} ,  \omega_{\nu} ) ,	
	\label{eq0prop0z} \\ &&
		\mathcal{D}_\sigma(\mathbf{q} , \omega_{\nu} )
	= 
		\frac	{- g c}
			{ 2 ( \omega_{\nu}^2 + \omega_{\bf q}^2 ) } 
	-
		\beta \mathcal{N} n_0
		\delta_{ \sigma , \uparrow }
		\delta_{ \omega_\nu , 0 }
		\delta_{ \mathbf{q}  , \mathbf{0} }, \\ && 
         \omega_{\bf q}=\sqrt{c^2{\bf q}^2+m^2} ,
	\label{eq0d}
\end{eqnarray} 
where ${\cal N}$ is the number of lattice sites. 
The saddle-point equation for the Lagrange multiplier $m^2$ reads
\begin{eqnarray}
		\frac{1}{\beta}
		\sum_{\omega_\nu}
		\int_{ |\mathbf{q}| < \Lambda }
		\frac	{g c}
			{ \omega_{\nu}^2 + \omega_{\bf q}^2 }
	+
		n_0
	=
		1
	\label{eq0m} .
\end{eqnarray}	 
In Eqs.~(\ref{eq0prop0z}-\ref{eq0m}), we have allowed for a Bose condensation of the
Schwinger bosons in the mode $\mathbf{q}=\mathbf{0}$, with
$n_0 = \frac{1}{\mathcal{N} \beta} 
\left< \mathrm{z}^{\dagger}( \mathbf{q} = \mathbf{0} , \omega_\nu = 0 )
\mathrm{z}( \mathbf{q} = \mathbf{0} , \omega_\nu = 0 ) \right>$
the fraction of condensed bosons. Bose condensation signals the
appearance of AF 
long-range order: $\langle {\bf n}_{\bf r}\rangle=n_0 {\bf u}_z$. Knowing the
propagator of the $z$ field, one can then calculate 
the spin-spin correlation function using Eq.~(\ref{eq0n0z}). The AF correlation length
$\xi$ is related to the mass $m$ of the bosonic propagator
$\mathcal{D}$ {\it via} $m = c/2\xi$.\cite{Auerbach94} $m$ vanishes
whenever the fraction of the condensed bosons is finite.
\begin{figure}
	\epsfig{file=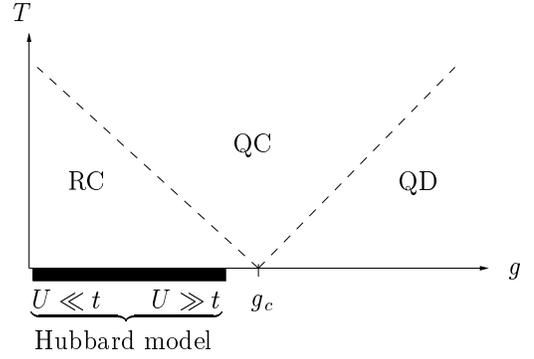}
	\caption{Phase diagram of the NL$\sigma$M derived from a
		saddle-point approximation 
		in the $\mathrm{CP}^1$ representation. At $T=0$, there
		is long-range order when the
		coupling constant $g < g_c = 4 \pi / \Lambda$. The
		three finite-temperature 
		regimes correspond to "renormalized classical" (RC),
		"quantum critical" (QC) 
		and "quantum disordered"
		(QD).\cite{Chakravarty89,Sachdev99} The ground state of the
		2D half-filled Hubbard model is ordered for any value
		of the Coulomb  
		repulsion $U$. At finite temperature, there are strong
		AF fluctuations 
		with an exponentially large correlation length $\xi \gg 1$ (RC regime).
		}
\label{fig0msnl}
\end{figure}

At zero temperature, the solution of the saddle-point equation (\ref{eq0m}) shows that
the NL$\sigma$M is ordered at small $g$ ($m = 0$ and $n_0 > 0$)
and disordered by quantum fluctuations at large $g$ ($m > 0$ and $n_0 = 0$). 
The two regimes are  separated by a quantum-critical point at 
$g_c=4 \pi/\Lambda$. In the ordered phase ($g\leq g_c$), the fraction
of condensed bosons is $n_0 = 1 - g/g_c$.


The condition of zero-temperature long-range order is satisfied in
the NL$\sigma$M derived from the half-filled Hubbard model
(Fig.~\ref{fig0msnl}). For $U\ll 4t$, $g/g_c \sim
e^{-2\pi\sqrt{t/U}}$ is exponentially small. For $U\gg 4t$,
$\rho_s^0\simeq J/4$ and $c\Lambda\simeq \sqrt{2}J$, so that
$g/g_c \simeq \sqrt{2}/\pi<1$. Notice that
setting the cutoff to a higher value at strong coupling would lead us into
the quantum disordered regime. However, our choice is consistent with
results obtained 
by mapping the Hubbard model at strong coupling onto the Heisenberg model.
It is known, both from numerical and analytical work, that the 2D
quantum Heisenberg model  on a square lattice is ordered at zero
temperature.\cite{Manousakis91}  

Fig.~\ref{fig0coefs} shows the fraction of condensed bosons as a function of $U$.
For this, and subsequent, numerical calculations we use a smooth cutoff, i.e.
$ \int_{ |\mathbf{q}| < \Lambda } \rightarrow \int_{\mathbf{q}} 
\frac{e^{-|\mathbf{q}| \xi_0} - e^{-q_0 \xi_0}}{1 -  e^{-q_0 \xi_0}}$.
In contrast to a hard cutoff, this procedure prevents artificial features in
the fermion spectral function and in the density of states.
The parameter $q_0$ is adjusted so as to reproduce in the strong-coupling limit
($U \gg 4t$) the result $|\left< \mathbf{n}_\mathbf{r} \right>| = n_0
\simeq 0.6$ obtained from the Heisenberg model.\cite{Manousakis91}
While the value of $n_0$ for $U\ll 4t$ and $U\gg 4t$ does not depend on
$\xi_0$, the behavior at intermediate coupling is
strongly cutoff dependent.   

At finite temperature, the AF long-range order is suppressed ($n_0=0, m>0$),
in agreement with the Mermin-Wagner theorem. For systems that exhibit
AF long-range 
order at $T=0$, the correlation length remains nevertheless exponentially large
at low temperature [renormalized classical regime, see
  Fig.~(\ref{fig0msnl})]. From Eq.~(\ref{eq0m}), we deduce
\begin{equation}
	\xi = \frac{c}{2m} 
	, \qquad
	m = T e^{  
		- \frac{2 \pi \rho_s}{T} 
		} ,
	\label{eq0m0expo}
\end{equation} 
where $\rho_s=\rho_s^0(1-g/g_c)$ is the zero-temperature spin stiffness. 
The mass $m$ of the bosonic propagator being much smaller than the
temperature, the dominant fluctuations are classical. 

\begin{figure}
	\epsfig{file=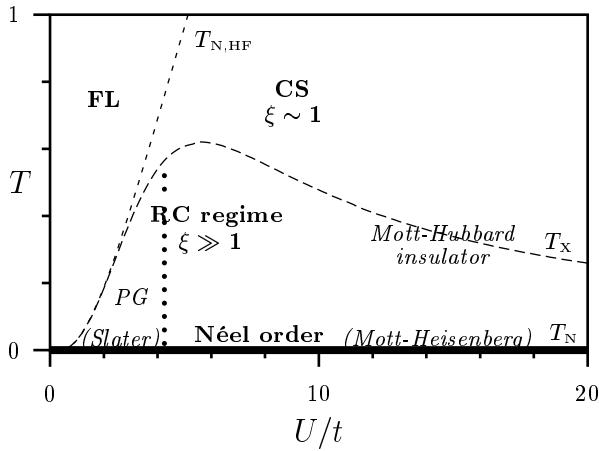}
	\caption{Phase diagram of the 2D half-filled Hubbard model.
		$T \gtrsim T_{\mathrm{N}}^{\mathrm{HF}}$: Fermi liquid
		(FL) phase; 
		$T_\mathrm{X} \lesssim T \lesssim
		T_{\mathrm{N}}^{\mathrm{HF}}$: local moments 
		with no AF short-range order (Curie spins, $\xi \sim 1$);
		$T=0$: Slater ($U \ll 4t$) and Mott-Heisenberg ($U \gg 4t$) 
		antiferromagnets. At finite temperature, there is a
		pseudogap phase  
		($U \ll 4t$) and a Mott-Hubbard insulator ($U \gg 4t$)
		separated by a  
		metal-insulator transition (dotted line) defined by
		the vanishing of the tunneling density of states
		$\rho(\omega=0)$ at zero energy 
		(Sec.~\ref{sec0fc0spec}).
		All lines, except $T_{\rm N}=0$ (thick solid line),
		are crossover lines. The 
		NL$\sigma$M description is valid below $T_X$ (RC regime). 
		[From Ref.~\onlinecite{KBND03}.]
		}
\label{fig0diag0phases0magentique}
\end{figure}

Let us now discuss the limits of validity of the NL$\sigma$M. The
derivation of the NL$\sigma$M is based on the assumption that the
dominant low-energy fluctuations 
are transverse spin waves with a large correlation length. The condition
$T \ll T_{\mathrm{N}}^{\mathrm{HF}}$ ensures that amplitude
fluctuations of the AF order parameter are frozen at low energy.
One should also verify that the computation of $\xi$ within the NL$\sigma$M
is consistent with the assumption of AF short-range order, i.e. $\xi
\gg \Lambda^{-1}$ or, equivalently, $m \ll
c \Lambda/2$. We define $T'$ as the solution of the equation $m \sim
c\Lambda/2$ obtained from Eq.~(\ref{eq0m}). Then, the domain of
validity of the NL$\sigma$M is given by 
$T \ll T_\mathrm{X} \sim \mathrm{min} ( T_{\mathrm{N}}^{\mathrm{HF}} ,
T' )$. At weak coupling, $T_X\sim T_{\rm N}^{\rm HF}$, while $T_X \sim
T'\sim J$ at strong coupling.  The crossover 
temperature $T_\mathrm{X}$ displayed in Fig.~\ref{fig0diag0phases0magentique}
is a smooth interpolation between  $T_{\mathrm{N}}^{\mathrm{HF}}$ and
$T'$.\cite{note0interp}

The phase diagram is shown in Fig.~\ref{fig0diag0phases0magentique}. 
Above $T_{\mathrm{N}}^{\mathrm{HF}}$, spin fluctuations are not
important and we expect  
a Fermi liquid behavior. Between $T_{\mathrm{N}}^{\mathrm{HF}}$ and
$T_\mathrm{X}$ (a regime which exists only in the strong-coupling limit), local
moments form but with no AF short-range order (Curie spins: $\xi \sim
1$). Below $T_\mathrm{X}$, the system 
enters a renormalized classical regime of spin fluctuations where the AF
correlation length becomes exponentially large [Eq.~(\ref{eq0m0expo})]. AF
long-range order sets in at $T_\mathrm{N} = 0$. Although there is a smooth
evolution of the magnetic properties as a function of $U$, the physics is quite
different for $U \ll 4t$ and $U \gg 4t$. This will be shown in
Sec.~\ref{sec0fc0spec} by studying the fermion spectral properties. The main
conclusions are shown in Fig.~\ref{fig0diag0phases0magentique}. At zero
temperature the system is an antiferromagnet, which evolves from a
Slater to a Mott-Heisenberg behavior as $U$ increases. At finite temperature
there is a pseudogap phase for $U \ll 4t$ and a Mott-Hubbard insulator
for $U \gg 4t$. These two regimes are separated by a
(finite-temperature) metal-insulator transition (dotted line in 
Fig.~\ref{fig0diag0phases0magentique}) defined by
		the vanishing of the tunneling density of states
		$\rho(\omega=0)$ at zero energy. 

\section{Fermion spectral properties}
\label{sec0fc0spec}

In this section, we study the influence of the long-wavelength spin
fluctuations on the fermion spectral properties. The fermionic Green's
function $- \left< \Psi_{\mathbf{r}_1 \tau_1} \Psi_{\mathbf{r}_2
  \tau_2}^{\dagger} \right>$,  
written here as a $2 \times 2$ matrix in spin space,
can easily be related to the pseudo-fermions by use of the relation
$\Psi_\mathbf{r} = R_\mathbf{r} \Phi_\mathbf{r}$:
\begin{equation}
		\mathcal{G}(1,2)
	=	
	- \left<	
		R_{1} 
		\Phi_{1} 
		\Phi_{2}^{\dagger} 
		R_{2}^{\dagger} 
	\right> .
	\label{eq0green0r0phi}
\end{equation}
Here we use the shorthand notations
$1\equiv (\mathbf{r}_1,\tau_1)$ and $2\equiv (\mathbf{r}_2,\tau_2)$.
The averaging in the above expression should be performed with respect to
the action 
$S_{\mathrm{HF}}[ \Phi ] + S'[ \mathrm{z} , \Phi , \mathbf{L}]$ 
obtained in Sec.~\ref{sec0fluctuations0spin} from the second-order expansion 
in $\mathbf{L}$
and $\partial_\mu\mathbf{n}$. $S'$ stands for the sum of the perturbative
corrections 
$S_{\mathrm{p}}$, $S_{\mathrm{d}}$, $S_{\mathrm{l}}$, $S_{\mathrm{l^2}}$
defined in (\ref{eq0act0para}-\ref{eq0act0l02}). 
Integrating first the pseudo-fermions, we can write the propagator as
\begin{eqnarray}
		\mathcal{G}(1,2)
	&=&
		\frac{1}{Z}
		\int \mathcal{D}[\mathrm{z}]
		e^{ -S_{\mathrm{NL}\sigma\mathrm{M}} [ \mathrm{z} ] }
		R_1
		\mathcal{G}(1,2|\mathrm{z})
		R_2^{\dagger} 
	 , \\
		Z
	&=&
      \int\mathcal{D}[\mathrm{z}] e^{ -S_{\mathrm{NL}\sigma\mathrm{M}}[z] },   
	\label{eq0green0comme0moy0z}
\end{eqnarray}
where $\mathcal{G}(1,2|\mathrm{z})$ is the pseudo-fermion propagator
calculated for a given configuration of the bosonic field $\mathrm{z}$:
\begin{equation}
		\mathcal{G}(1,2|\mathrm{z})
	=
		-
		\frac	{ 
			\int \mathcal{D}[ \Phi , \mathbf{L}] 
			\phi_1 \phi_2^{\star}
			e^	{
				- S_{ \mathrm{HF} }[\Phi] 
				- S'[ \mathrm{z} , \Phi , \mathbf{L} ] 
				}
			}
			{ 
			\int \mathcal{D}[ \mathrm{ \Phi , \mathbf{L} } ] 
				e^	{
					- S_{ \mathrm{HF} }[\Phi] 
					- S'[ \mathrm{z} , \Phi , \mathbf{L} ] 
					}
			}
	\label{eq0green0pseudo0fermion0pour0z0donne} .
\end{equation}	 
The action 
$S_{\mathrm{HF}}[\Phi] + S'[\mathrm{z},\Phi,\mathbf{L}]$ 
describes HF fermions  interacting with spin fluctuations {\it via}
the action $S'$. Since the  
HF pseudo-fermions are gapped, we expect a perturbative expansion in
$S'$ to be well-behaved. To leading order,
$\mathcal{G}(1,2|\mathrm{z})=\mathcal{G}^{\mathrm{HF}}(1,2)$ and the
fermion Green's function simplifies into 
\begin{eqnarray}
		\mathcal{G}_{ \sigma_1 \sigma_2 } (1,2)
	&=&
		\sum_{ \alpha_1, \alpha_2 }
		\mathcal{G}_{ \alpha_1 \alpha_2 }^{\mathrm{HF}}( 1 , 2 )
		\left<
			(R_1)_{ \sigma_1 \alpha_1 } 
			(R_2)_{ \sigma_2 \alpha_2 }^{\star}
		\right>	
	\nonumber , \\
	\label{eq0decouplage}
\end{eqnarray}
where the product of rotation matrices is averaged with the
NL$\sigma$M action. This approximation neglects the effect of spin
fluctuations on the propagation of pseudo-fermions. Their  
influence on the propagation of fermions is implemented only through the 
decomposition of the fermion into a boson and a pseudo-fermion.

Using the Schwinger boson propagator derived in Sec.~\ref{sec0modele0sigma}
[Eqs.~(\ref{eq0prop0z}-\ref{eq0d})], we have
\begin{eqnarray}
		\left<
			(R_1)_{ \sigma_1 \alpha_1 } 
			(R_2)_{ \sigma_2 \alpha_2 }^{\star}
		\right>
	&=&
		-
		\delta_{ \sigma_1 , \sigma_2} \delta_{ \alpha_1 , \alpha_2}
		\nonumber \\ && \times 
                    \left(
			\overline{\mathcal{D}}(1,2)
			-
			\delta_{ \sigma_1 , \alpha_1 } 
			n_0
		\right) ,
	\label{eq0moyennes0r}
\end{eqnarray}
where $\overline{\mathcal{D}}$ is the non-condensed part of $\mathcal{D}_\sigma$. Using this 
expression in Eq.~(\ref{eq0decouplage}) we finally obtain for the fermion Green's function:
\begin{widetext}
\begin{eqnarray}
		- \left< 
			\psi_{\mathbf{r} \tau \sigma}
			\psi_{\mathbf{r'} \tau' \sigma'}^{\star}
		\right> 
	&=&
		\delta_{\sigma , \sigma'} 
		\mathcal{G}_{\sigma}(\mathbf{r} , \mathbf{r}' , \tau - \tau' ) 
	\label{eq0prop0phys} , \\
		\mathcal{G}_{\sigma} ( \mathbf{k} , \mathbf{k}' , \omega ) 
	&=&
		- \frac	{ 
			2 \delta_{ \mathbf{k} , \mathbf{k}' }
			}
			{
			\beta
			}
		\sum_{\omega_{\nu}}
		\int_{ \mathbf{q} }
		\mathcal{G}_{\sigma}^{\mathrm{HF}}
	          ( \mathbf{k-q} , \mathbf{k-q} , \omega - \omega_{\nu} ) 
		 \overline{\mathcal{D}}( \mathbf{q} , \omega_{\nu} )
	+
		n_0 
		\mathcal{G}_{\sigma}^{\mathrm{HF}}( \mathbf{k} ,
	\mathbf{k}' ,  \omega )  .
	\label{eq0g0phys0four}
\end{eqnarray}
Since $n_0$ vanishes at finite temperature, the fermion Green's
function is spin-rotation and translation invariant in the absence of
AF long-range order. We show below that the first term of the rhs of
(\ref{eq0g0phys0four}) corresponds to incoherent excitations. At zero
temperature, the last term of (\ref{eq0g0phys0four}) describes
Bogoliubov QP's carrying a total spectral weight $n_0$. 

To study in detail the fermion excitations, we consider the spectral function
$\mathcal{A} ( \mathbf{k} , \omega ) = - \pi^{-1} \mathrm{Im} 
\mathcal{G}_{\sigma} ( \mathbf{k} , \mathbf{k} , i \omega \rightarrow
\omega + i 0^+ )$ and the tunneling density of states (DOS)
$\rho(\omega)=\int d \omega \mathcal{A}( \mathbf{k}, \omega )$.
Performing the summation over bosonic Matsubara frequencies in
Eq.~(\ref{eq0g0phys0four}) we obtain
\begin{eqnarray}
		\mathcal{A} ( \mathbf{k} , \omega )
	&=&
		\mathcal{A}_{\mathrm{inc}} ( \mathbf{k} , \omega )
		+ n_0 \mathcal{A}_{\mathrm{HF}} ( \mathbf{k} , \omega )
	\label{eq0fc0spec0decomp} , \\
		\mathcal{A}_{\mathrm{inc}} ( \mathbf{k} , \omega )
	&=&
		\int_{\mathbf{q}} \frac{g c}{2 \omega_q}
		\Bigl\lbrace
                 \left[
			n_B( \omega_\mathbf{q} ) + n_F( -E_{\mathbf{k-q}} )
		\right]
		\left[
			u_{\mathbf{k-q}}^2 
			\delta( \omega - \omega_{\mathbf{q}} - E_{\mathbf{k-q}} )
			+
			v_{\mathbf{k-q}}^2 
			\delta( \omega + \omega_{\mathbf{q}} + E_{\mathbf{k-q}} )
		\right]
	\nonumber \\
	&& +
		\left[
			n_B( \omega_\mathbf{q} ) + n_F( E_{\mathbf{k-q}} )
		\right]
		\left[
			u_{\mathbf{k-q}}^2 
			\delta( \omega + \omega_{\mathbf{q}} - E_{\mathbf{k-q}} )
			+
			v_{\mathbf{k-q}}^2 
			\delta( \omega - \omega_{\mathbf{q}} + E_{\mathbf{k-q}} )
		\right] \Bigr\rbrace  ,
	\label{eq0fc0spec0inc} 
\end{eqnarray}
\end{widetext}
where $n_F(\omega)$ and $n_B(\omega)$ are the usual Fermi and Bose occupation 
numbers $(e^{\beta \omega} \pm 1)^{-1}$, and
$\mathcal{A}_{\mathrm{HF}}$ the HF spectral function: 
\begin{eqnarray}
		\mathcal{A}_{ \mathrm{HF} } ( \mathbf{k} , \omega )	
	=
		u_{\mathbf{k}}^2
		\delta( \omega - E_{\mathbf{k}} )
		+
		v_{\mathbf{k}}^2
		\delta( \omega + E_{\mathbf{k}} )
	\label{eq0fc0spec0hf} , \\
		u_{\mathbf{k}}^2
	=
		\frac{1}{2}
		\left(
			1 + \frac{\epsilon_\mathbf{k}}{E_\mathbf{k}}
		\right)
	, \quad
		v_{\mathbf{k}}^2
	=
		\frac{1}{2}
		\left(
			1 - \frac{\epsilon_\mathbf{k}}{E_\mathbf{k}}
		\right)
	\label{eq0u0v} .
\end{eqnarray}	 
One can check that the spectral function
$\mathcal{A}( \mathbf{k} , \omega )$
is normalized to unity. From
Eqs.~(\ref{eq0fc0spec0decomp}-\ref{eq0fc0spec0inc}) we deduce
\begin{eqnarray}
		\int d \omega \mathcal{A}( \mathbf{k} , \omega ) 
	&=&
		\int_{|\mathbf{q}| < \Lambda}
		\frac{g c}{\omega_\mathbf{q}}
		\left( 
			n_B(\omega_{\bf q}) + \frac{1}{2}
		\right)
		+ n_0 \nonumber \\ 
	&=&
		1 ,
	\label{eq0normalisation0de0fc0spec}
\end{eqnarray} 
where the second equality is obtained by using
$\left< \mathrm{z}_{\mathbf{r}}^{\dagger}  \mathrm{z}_\mathbf{r} \right>
= 1$ [Eq.~(\ref{eq0m})].
From Eqs.~(\ref{eq0fc0spec0decomp}-\ref{eq0u0v}), we obtain
\begin{eqnarray}
		\rho( \omega ) 
	&=&	
		\rho_{\mathrm{inc}}( \omega ) 
	+
		n_0 \rho_{\mathrm{HF}}( \omega ) 
	\label{eq0dos0decomp} , \\
		\rho_{\mathrm{inc}}( \omega ) 
	&=&	
		\rho_{\mathrm{inc}}^>( \omega ) 
	+
		\rho_{\mathrm{inc}}^>( - \omega ) 
	\label{eq0dos0decomp02} , \\
		\rho_{\mathrm{inc}}^>( \omega ) 
	&=& \frac{g}{4\pi c} 
		\int_{m}^{c \Lambda} d \omega' 
		\Big[
			n_B(\omega') 
			\rho_{\mathrm{HF}}( \omega + \omega' )
			\theta( \omega + \omega' )
	\nonumber \\
	&& + \;
			( n_B(\omega') + 1 )
			\rho_{\mathrm{HF}}( \omega - \omega' )
			\theta( \omega - \omega' )
		\Big]
	\label{eq0dos} ,
\end{eqnarray}
where
\begin{equation}
		\rho_{\mathrm{HF}}( \omega )
	=
		\theta( \omega^2 - \Delta_0^2 )
		\frac{ |\omega| }{ \sqrt{\omega^2-\Delta_0^2} }
		\;
		\rho_0 \left( \sqrt{\omega^2-\Delta_0^2} \right)
	\label{eq0dos0hf}
\end{equation}
is the HF DOS and $\theta$ the step function. We have approximated 
the Fermi occupation numbers by their zero-temperature limit, which is
valid for $T \ll T_{\mathrm{N}}^{\mathrm{HF}}$.

\subsection{$T=0$: Slater {\it vs} Mott-Heisenberg antiferromagnetism} 
\label{sec0t0nulle}

At zero temperature, the incoherent part of the spectral function 
[Eq.~(\ref{eq0fc0spec0inc})] can be simplified.
All the occupation factors vanish, except fermionic factors at
negative energies which are equal to 1, so that
\begin{eqnarray}
		\mathcal{A}_{ \mathrm{inc} } ( \mathbf{k} , \omega ) 
	=
		\int_{\mathbf{q}}
		\frac{g c}{2 \omega_q}
		[
			u_{\mathbf{k-q}}^2 
			\delta( \omega - \omega_{\mathbf{q}} - E_{\mathbf{k-q}} )
	\nonumber \\
			+
			v_{\mathbf{k-q}}^2 
			\delta( 
				\omega + 
				\omega_{\mathbf{q}} + 
				E_{ \mathbf{k} - \mathbf{q} }
			)
		]
	\label{eq0fc0spec0zero} .
\end{eqnarray}
In the same way, we obtain for the DOS
\begin{equation}
		\rho_{\mathrm{inc}}^>( \omega ) 
	=
		\frac{g}{4 \pi c}
		\int_{0}^{c \Lambda} d \omega'
		\rho_{\mathrm{HF}}( \omega - \omega' )
		\theta( \omega - \omega' )
	\label{eq0dos0zero} .
\end{equation}

\begin{figure}
	\epsfig{file=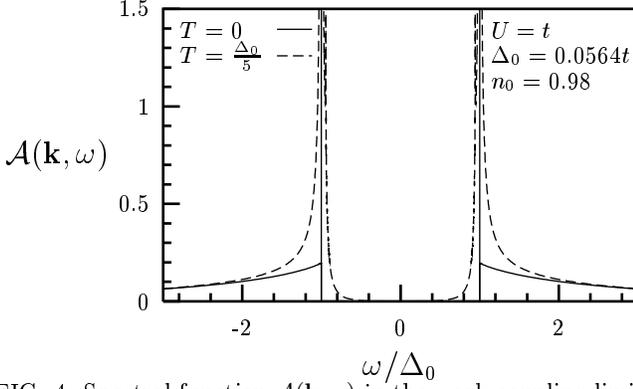}
	\caption{Spectral function $\mathcal{A}(
	\mathbf{k}, \omega)$ in the weak-coupling limit $U = t$ for
	$T=0$ (Slater antiferromagnet) and $T=\Delta_0/5$ (pseudogap
	phase). $\mathbf{k} = (\pi/2,\pi/2)$. The vertical
	lines represent Dirac peaks of weight $n_0/2$ (Bogoliubov
	QP's). At finite temperature, precursors of the
	zero-temperature Bogoliubov QP's show up as peaks of width
	$\sim  T$ at $\pm E_{\bf k}$. At low energy (and $T>0$), we
	observe a pseudogap with an exponentially small spectral
	weight at $\omega=0$. 
        Energies are measured in units of $t$. 
	[From Ref.~\onlinecite{KBND03}.]
	}
	\label{fig0fc0spec0tnulle0ufaible}
\end{figure}

\medskip
	
\begin{figure}
	\epsfig{file=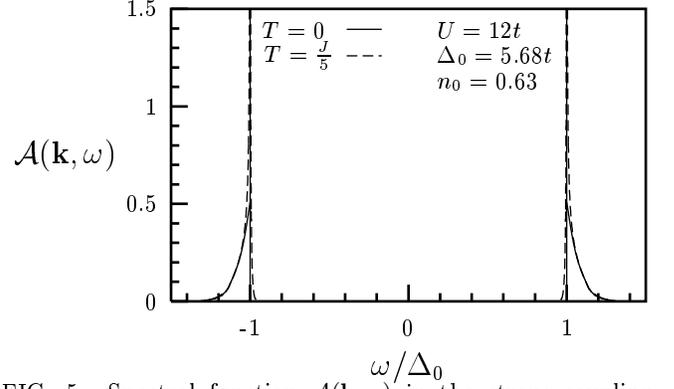}
	\caption{Spectral function ${\cal A}({\bf k},\omega)$ in the 
	strong-coupling regime $U = 12 t$ for $T=0$ (Mott-Heisenberg
	antiferromagnet) and $T=J/5$ (Mott-Hubbard insulator).
	At $T=0$, when $U$ increases, spectral weight is transferred from the
	Bogoliubov QP peaks to the incoherent excitation background.
	[Note the difference in the energy scale, which is fixed by
	$\Delta_0$, between Figs.~\ref{fig0fc0spec0tnulle0ufaible} and
	\ref{fig0fc0spec0tnulle0ufort}.] [From Ref.~\onlinecite{KBND03}.]
	}
\label{fig0fc0spec0tnulle0ufort}
\end{figure}

In Figs.~\ref{fig0fc0spec0tnulle0ufaible}-\ref{fig0fc0spec0tnulle0ufort}
we show the spectral function at the 
$\mathbf{k} = ( \pi/2 , \pi/2 )$ 
point of the non-interacting Fermi surface at weak $(U=t)$ and strong  
$(U = 12 t)$ coupling. The spectral function
$\mathcal{A}( \mathbf{k} , \omega )$
exhibits a gap $2\Delta_0$, which is a consequence of AF long-range
order. There are well-defined Bogoliubov QP's with excitation energy
$\pm E_\mathbf{k}$, as in HF theory, 
but their spectral weight is reduced by a factor $n_0 < 1$ because of 
quantum spin  
fluctuations. The remaining weight $(1-n_0)$ is carried by an
incoherent excitation background at higher energy
($|\omega|>E_\mathbf{k}$).  

There are important differences between the weak $(U \ll 4t)$ and
strong $(U \gg 4t)$ 
coupling regimes. First, the AF gap $2\Delta_0 \sim t e^{-2 \pi \sqrt{t/U}}$ is
exponentially small at weak coupling, while it tends to $U$ for $U \gg 4t$.
Second, the Bogoliubov QP's carry most of the spectral weight in the
weak-coupling 
regime, since $g/g_c=1-n_0$ is exponentially small when $U \ll 4t$. As $U$
increases, spectral weight is transferred from the Bogoliubov QP's to
the incoherent excitation 
background, and at strong coupling $( U \gg 4t)$ the incoherent
excitation background carries
a significant fraction of the total spectral weight (i.e. $n_0$ and $1-n_0$ are of
the same order). Third, the energy range of the incoherent excitation background depends on
the value of $U$. From Eq.~(\ref{eq0fc0spec0zero}) we see that it extends from 
$E_\mathbf{k}$ to $\sim \sqrt{ E_\mathbf{k}^2 + 16 t^2 \Lambda^2 } + c \Lambda$.
At weak coupling, the upper limit 
turns out to be of order $\Delta_0$ (for $\mathbf{k}$ lying on the non-interacting
Fermi surface). Thus, the energy range of the incoherent excitation background remains very
small with respect to the dispersion of the Bogoliubov QP energy
$E_\mathbf{k}$, which is of order $t$ when $\Delta_0 \ll t$. 
At strong coupling, the incoherent excitation background above $E_{\bf k} \sim
U/2$ extends over a range of order $J$. This 
energy range is of the same order of magnitude as the dispersion of
the Bogoliubov QP energy, which is also of order $J$ when $U \gg 4t$
[as can be seen from the expansion 
$E_\mathbf{k} \simeq U/2 + J(\cos{k_x}+\cos{k_y})^2$].

\begin{figure}
	\epsfig{file=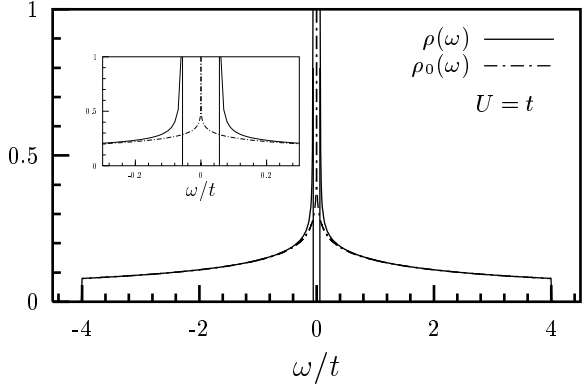}
	\caption{Zero-temperature DOS $\rho(\omega)$ at weak coupling:
		$U = t$ (Slater antiferromagnet). $\rho(\omega)$
		differs from the free-fermion 
		DOS $\rho_0(\omega)$ only at low energy due to the
		opening of the AF gap $2\Delta_0$ (see inset). Since
		the incoherent excitation background carries a
		negligible fraction of the total spectral weight,
		there is no noticeable difference between
		$\rho(\omega)$ and the HF DOS $\rho_{\rm
		HF}(\omega)$ (not shown in the figure). 
		}
\label{fig0dos0tzero0ufaible}
\end{figure}

\begin{figure}
	\epsfig{file=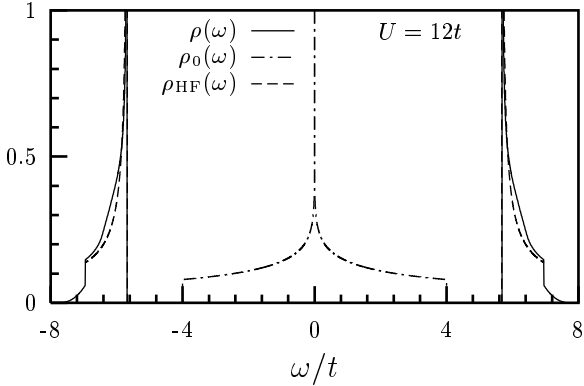}
	\caption{Same as Fig.~\ref{fig0dos0tzero0ufaible}, but at strong
		coupling: $U = 12 t$ (Mott-Heisenberg antiferromagnet). 
		$\rho(\omega)$ differs strongly from the non-interacting DOS
		$\rho_0(\omega)$, as the AF gap exceeds the non-interacting
		bandwidth. It also differs from the
		HF DOS $\rho_\mathrm{HF}(\omega)$ due to the
		incoherent excitation 
		background carrying a significant fraction of the
		total spectral weight.
		}
\label{fig0dos0tzero0ufort}
\end{figure}

In Figs.~\ref{fig0dos0tzero0ufaible}-\ref{fig0dos0tzero0ufort} we compare
the zero-temperature DOS $\rho(\omega)$  and the non-interacting DOS
$\rho_0(\omega)$. 
At weak coupling ($U = t$), $\rho(\omega)$ is similar to the HF
result, with no visible effect of the incoherent excitation background.
$\rho(\omega)$ differs from $\rho_0(\omega)$ mainly at low energy, due
to the (small) AF gap $2\Delta_0$. At strong coupling ($U = 12 t$),
$\rho(\omega)$ 
differs strongly from $\rho_0(\omega)$, due to an AF gap $2\Delta_0
\sim U$ exceeding 
the non-interacting bandwidth. There is also a significant difference
between $\rho(\omega)$ and $\rho_\mathrm{HF}(\omega)$, which results from
the incoherent excitation background.

The spectral function $\mathcal{A} ( \mathbf{k} , \omega )$ and the DOS
$\rho ( \omega )$ are typical of a Slater antiferromagnet at weak
coupling and of a 
Mott-Heisenberg antiferromagnet at strong coupling. As shown in the
next section, 
Slater and Mott-Heisenberg antiferromagnets behave very differently at finite
temperature.

\subsection{$T > 0$: pseudogap {\it vs} Mott-Hubbard gap}
\label{sec0t0finie}

At finite temperature, $n_0$ vanishes and $\mathcal{A} =
\mathcal{A}_{\mathrm{inc}}$.  
The result of the numerical calculation for $U=t$ and $U = 12 t$  for
$\mathbf{k} = ( \pi/2 , \pi/2 )$ is shown in
Figs.~\ref{fig0fc0spec0tnulle0ufaible} 
and \ref{fig0fc0spec0tnulle0ufort}. $\mathcal{A} ( \mathbf{k} , \omega
)$ exhibits broadened peaks of width $T$ at the HF QP 
energy $\pm E_\mathbf{k}$. These peaks are incoherent precursors of
the zero-temperature Bogoliubov QP peaks. The zero-temperature AF gap
is partially filled at strong coupling and transforms into a  
pseudogap in the weak-coupling regime. At higher energy
($|\omega|\gtrsim E_{\bf k}$), a roughly
featureless incoherent excitation background is observed.

\subsubsection{Precursors of Bogoliubov QP's}
\label{sec0qp}

At finite temperature, the coherent part of the spectral function disappears.
However, sharp peaks are still observed at the HF energy $\pm E_\mathbf{k}$.
To study the peak at $E_{\bf k}$, let us perform a few approximations on the
finite-temperature 
spectral function (\ref{eq0fc0spec0inc}). First, at positive energies,
almost all the spectral weight comes from the terms proportional to
$u^2_{{\bf k}-{\bf q}}$ in (\ref{eq0fc0spec0inc}) (except 
at energies close to zero), whose sum will be denoted by
$\mathcal{A}^>$. Second, we replace the Fermi occupation number by the
step function, given that the temperature is small  
compared to $E_\mathbf{k}$. Regrouping terms containing the Bose occupation 
numbers we obtain
\begin{eqnarray}
		\mathcal{A}^> ( \mathbf{k} , \omega ) 
	&=&
		\mathcal{A}_{\mathrm{bg}}^> ( \mathbf{k} , \omega ) 
	+
		\mathcal{A}_{\mathrm{peak}}^> ( \mathbf{k} , \omega ) 
	\label{eq0fc0spec0pic0fond} , \\		
		\mathcal{A}_{\mathrm{peak}}^> ( \mathbf{k} , \omega )
	&=&
		\int_{\mathbf{q}}
		\frac{g c}{2 \omega_\mathbf{q}}
		n_B( \omega_\mathbf{q} )
		u_{\mathbf{k-q}}^2 
		\big[
			\delta( \omega - \omega_{\mathbf{q}} - E_{\mathbf{k-q}} )
	\nonumber \\
	&&	+
			\delta( \omega + \omega_{\mathbf{q}} - E_{\mathbf{k-q}} )
		\big]
	\label{eq0fc0spec0pic} .
\end{eqnarray}
$\mathcal{A}_{\mathrm{bg}}^>$ has the same expression as the
incoherent excitation background term 
(\ref{eq0fc0spec0zero}) at zero temperature. It thus describes a 
temperature-independent incoherent excitation background at energies
above $E_{\bf k}$.
$\mathcal{A}_{\mathrm{peak}}^>$ gives rise to the peak at the HF
energy $E_{\bf k}$. To see this, let us
put it into a more explicit form. Because of the bosonic occupation numbers,
the sum over ${\bf q}$ in (\ref{eq0fc0spec0pic}) is dominated by wave-vectors
satisfying $\omega_{\mathbf{q}} \lesssim T$ or, equivalently, $|{\bf q}|
\lesssim T/c$. For $T\ll T_X$, $T/c\ll 1$ and we can neglect the
$\mathbf{q}$ dependence  
of $E_{\mathbf{k} - \mathbf{q}}$ and $u_{\mathbf{k}-\mathbf{q}}^2$.
The integrand then becomes isotropic, and one can use
$\int_{\mathbf{q}} \frac{ c^2 }{ \omega_{\mathbf{q}} } =
\int_{m}^{c \Lambda} \frac{d \omega_{\mathbf{q}} }{2 \pi}$.  
The result is 
\begin{equation}
	\mathcal{A}^>_{\mathrm{peak}} ( \mathbf{k} , \omega) =u_\mathbf{k}^2
	\frac{g}{4 \pi c} n_B( | \omega - E_\mathbf{k} | )
	\label{eq0fc0spec0pic0approx}
\end{equation}
for $|\omega - E_\mathbf{k}| > m$, and vanishes for $|\omega-E_\mathbf{k}|<m$.
For $m<| \omega - E_\mathbf{k} | \ll T$, $\mathcal{A}^>_{\mathrm{peak}}
( \mathbf{k} , \omega)$ behaves like 
$T/|\omega - E_\mathbf{k}|$. At energies further away from the
peak center,  
it decreases like $e^{-|\omega - E_\mathbf{k}|/T}$. Thus the
width of the peak is of the order of the temperature and therefore
corresponds to incoherent excitations. The vanishing of
$\mathcal{A}(\mathbf{k},\omega)$ for $|\omega-E_{\bf k}|<m$ is clearly
unphysical (note that 
it cannot be seen in the figures, since $m$ is exponentially
small). It would be suppressed by any finite life time in the bosonic
propagator ${\cal D}$. The finite-temperature DOS suffers from the
same artifact (i.e. $\rho(\omega)=0$ for $|\omega-\Delta_0|<m$).

The spectral weight of the peak at $E_\mathbf{k}$ is
\begin{eqnarray}
		\int d \omega \mathcal{A}_{\mathrm{peak}}^{>} ( \mathbf{k} , \omega )
	&=&
		u_{\mathbf{k}}^2 \frac{g}{2 \pi c} T 
		\ln \left( \frac{T}{m} \right)
	\nonumber \\
	&=&
		u_{\mathbf{k}}^2 \left( 1 - \frac{g}{g_c} \right)
	\label{eq0poids0pic} .
\end{eqnarray}
where the last result is obtained using Eq.~(\ref{eq0m0expo}). The spectral
weight of the peak turns out to be temperature independent and equal to
$u_\mathbf{k}^2 n_0$ ($n_0=1-g/g_c)$, which is nothing else but the
Bogoliubov QP weight in the ground state. We conclude that the peak
is an incoherent precursor of the zero-temperature
Bogoliubov QP peak. As the temperature decreases, it
retains its spectral weight but becomes sharper
and sharper, and eventually becomes a Dirac peak at $T=0$. As
expected, the spectral function evolves continuously when $T\to 0$.  
As in the zero-temperature case, the dependence of $n_0$ upon $U$
describes the transfer of spectral weight from the Bogoliubov QP's to the
incoherent excitation background when the Coulomb repulsion
increases. 

\begin{figure}
	\epsfig{file=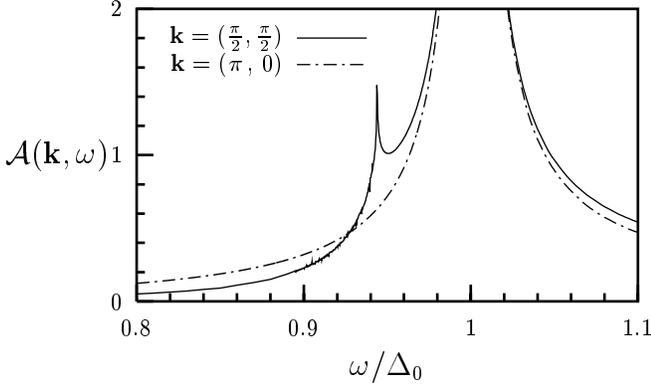}
	\caption{Finite-temperature spectral function at weak coupling
	for two different points of the non-interacting Fermi surface.
	For $\mathbf{k}$ close to ($\pi/2,\pi/2$)
	a second peak appears below $\Delta_0$ (see text).
	}
\label{eq0coefs0fig}
\end{figure}

The approximation (\ref{eq0fc0spec0pic0approx}) suggests that the peak
in ${\cal A}({\bf k},\omega)$ should exhibit the same
features, regardless of the location of ${\bf k}$ on the
non-interacting Fermi surface. Numerical calculations confirm 
this conclusion, with one exception. For wave-vectors near $( \pi/2 ,
\pi/2 )$, a second (smaller) peak appears at low  energy
(Fig.~\ref{eq0coefs0fig}). From a mathematical point of 
view, it is due to the vanishing of the first-order derivative of the
argument of the delta function in 
Eq.~(\ref{eq0fc0spec0pic}), which occurs for
$ \boldsymbol{\nabla}_{\bf q} \omega_\mathbf{q} =
\boldsymbol{\nabla}_{\bf q} E_{ \mathbf{k} - \mathbf{q} }$.
The energy at which the integration contour in the $\mathbf{q}$ plane,
defined by the delta  
function,  passes through this point can be estimated to be
$ \Delta_0 \sqrt{ 1 - ( c / |\mathbf{v}_\mathbf{k}| )^2 }$,
where
$\mathbf{v}_\mathbf{k}=\boldsymbol{\nabla}_{\bf k} \epsilon_{\bf k}$
is the free-fermion velocity. For wave-vectors verifying
$|\mathbf{v}_\mathbf{k}| < c$, i.e. sufficiently close to the Van-Hove
singularities, the second peak disappears. 
We believe this second peak to be an artifact of our lowest-order
approximation in the pseudo-fermion-boson interaction.

\subsubsection{Pseudogap {\it vs} Mott-Hubbard gap}
\label{subsubsec0t0finie}

\begin{figure}
	\epsfig{file=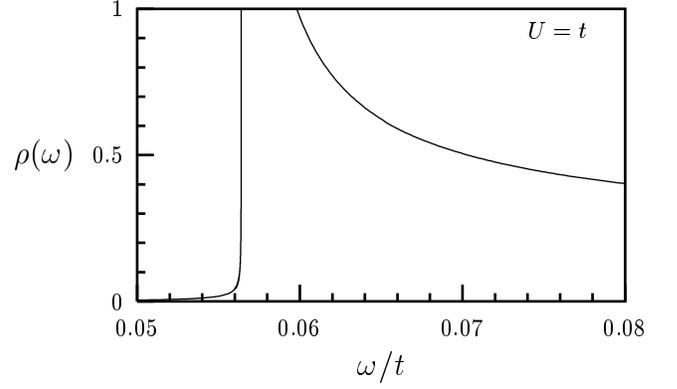}
	\caption{Finite-temperature DOS $\rho(\omega)$ at weak
		coupling: $U = t$, $T = \Delta_0/5$ (pseudogap phase).
		At $\omega = 0$ the DOS is finite but 
		exponentially small [Eq.~(\ref{dosw0})].
		}
\label{fig0dos0tfinie0ufaible}
\end{figure}

\begin{figure}
	\epsfig{file=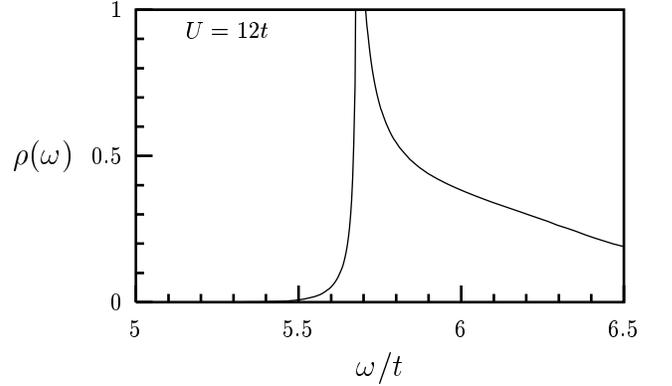}
	\caption{Finite-temperature DOS $\rho(\omega)$ at strong coupling:
		$U = 12 t$, $T = J/5$ (Mott-Hubbard insulator). 
	 	}
\label{fig0dos0tfinie0ufort}
\end{figure}

As shown in
Figs.~\ref{fig0fc0spec0tnulle0ufaible}-\ref{fig0fc0spec0tnulle0ufort}, the
spectral function $\mathcal{A} ( \mathbf{k} , \omega )$ extends below
the HF energy $E_\mathbf{k}$ (and above $-E_{\bf k}$ for $\omega<0$)
at finite temperature. The corresponding contribution to  
$\mathcal{A} ( \mathbf{k} , \omega )$ is given by [see
  Eq.~(\ref{eq0fc0spec0inc})] 
\begin{eqnarray}
		\int_\mathbf{q} \frac	{gc}
					{ 2 \omega_\mathbf{q}} 
		& n_B( \omega_\mathbf{q} ) &
		\Big[ 
			u^2_{\mathbf{k}-\mathbf{q}} 
			\delta( \omega + \omega_\mathbf{q} -
					E_{\mathbf{k}-\mathbf{q}} ) 
	\nonumber \\
			&& +  v^2_{\mathbf{k}-\mathbf{q}} 
			\delta( \omega - \omega_\mathbf{q} +
					E_{\mathbf{k}-\mathbf{q}} )  
		\Big] .
\end{eqnarray}
The presence of the Bose occupation number $n_B(\omega_\mathbf{q})$
shows that the 
low-energy fermion states ($|\omega| < E_\mathbf{k}$) are due to
thermal bosons,  
i.e. thermally excited spin fluctuations. A fermion added to the system
with momentum  
$\mathbf{k}$ and energy $|\omega| < E_\mathbf{k}$ can propagate by
absorbing a thermal  
boson of energy $\omega_\mathbf{q}$ and emitting a pseudo-fermion with energy 
$E_{\mathbf{k}-\mathbf{q}} = \omega + \omega_\mathbf{q}$. 

The lowest fermion energies are obtained by solving 
$\omega=E_{\mathbf{k}-\mathbf{q}}-\omega_\mathbf{q}$ 
(or  $\omega=-E_{\mathbf{k}+\mathbf{q}}+\omega_\mathbf{q}$). 
In the weak coupling limit, 
$\mathrm{max}_\mathbf{q} (\omega_\mathbf{q}) = c \Lambda \sim 2 \Delta_0$ 
and 
$E_{\mathbf{k}-\mathbf{q}} \sim E_\mathbf{k}$. 
Thus there is spectral weight at zero energy: the spectral function and the 
density of states exhibit a pseudogap 
(Figs.~\ref{fig0fc0spec0tnulle0ufaible} and \ref{fig0dos0tfinie0ufaible}). 
Note that the DOS remains exponentially small at low energy:
\begin{equation}
\rho(\omega) \sim e^{\frac{-\Delta_0}{T}} \cosh \left(
\frac{\omega}{T} \right) , \,\,\, 
|\omega| \ll \Delta_0. 
\label{dosw0}
\end{equation} 
This result differs from pseudogap theories based on
Gaussian spin fluctuations which find a much weaker suppression of the
density of states  
at low energy. \cite{Monien01} It bears some similarities with the
results obtained by  
Bartosch and Kopietz for fermions coupled to classical phase
fluctuations in incommensurate  
Peierls chains. \cite{Bartosch00} In the strong coupling limit,
thermally excited spin fluctuations lead to a small reduction of the
zero-temperature gap since $c\Lambda \sim J\ll E_{\bf k}\sim U/2$. The
system is a Mott-Hubbard insulator with a gap $2\Delta_0$ of order $U$
(Figs.~\ref{fig0fc0spec0tnulle0ufort} and \ref{fig0dos0tfinie0ufort}).

A last comment is in order here. Since the system is in the
renormalized classical regime,  
it is tempting to treat the NL$\sigma$M in the classical limit (which
amounts to  
neglecting the quantum (temporal) fluctuations of the N\'eel field
${\bf n}$). Such an  
approach is expected to be at least qualitatively correct for the
low-energy bosons  
($\omega_\mathbf{q} \lesssim T$) and should then give a good approximation of 
$\mathcal{A} (\mathbf{k},\omega)$ in the vicinity of the peaks around 
$\omega =\pm E_\mathbf{k}$. Retaining only the $\omega_\nu=0$ contribution in
Eq.~(\ref{eq0g0phys0four}), one finds 
\begin{eqnarray}
		\mathcal{A}_\mathrm{cl} (\mathbf{k},\omega) 
	= 	T \int_\mathbf{q}
		\frac{gc}{\omega^2_\mathbf{q}}
		\Big[ 
			u^2_{\mathbf{k}-\mathbf{q}} 
			\delta( \omega- E_{\mathbf{k}-\mathbf{q}} ) 
		&& \nonumber \\ 	+ 
			v^2_{\mathbf{k}-\mathbf{q}} 
			\delta(\omega+E_{\mathbf{k}-\mathbf{q}}) 
		\Big] . && 
\label{eq0fc0spec0classique}
\end{eqnarray}
Eq.~(\ref{eq0fc0spec0classique}) can also be obtained from Eq.~(\ref{eq0fc0spec0inc}) by 
using
$n_B(\omega_\mathbf{q}) + 1 \sim n_B(\omega_\mathbf{q}) \sim T/\omega_\mathbf{q} \gg 1$ 
and neglecting the term $\pm \omega_\mathbf{q}$ in the argument of the delta functions. 
It is readily seen that the classical calculation does not reproduce the pseudogap, 
since $\mathcal{A}_\mathrm{cl} ( \mathbf{k} , \omega )$ vanishes for $|\omega|<E_\mathbf{k}$. 
Although the pseudogap originates from thermally excited spin fluctuations in the 
renormalized classical regime, a fully quantum-mechanical calculation of 
$\mathcal{A} ( \mathbf{k} , \omega )$ turns out to be necessary to account for
the presence of low-energy fermion excitations.

\subsection{Finite-temperature metal-insulator transition} 
\label{subsec:mit}


\begin{figure}
        \centerline{\psfig{file=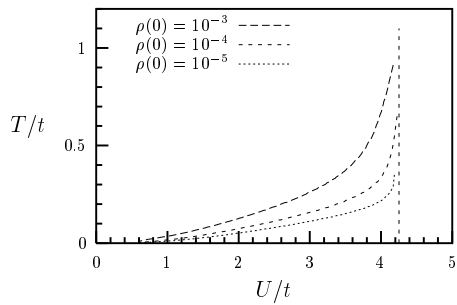,width=6.cm,angle=0}}
        \caption{Lines $\rho(\omega=0)={\rm const}$ in the $(U,T)$
        plane. The vertical line corresponds to $\rho(\omega=0)=0$. } 
\label{fig0dos1}
\end{figure}

We conclude from the results of Sec.~\ref{sec0t0finie} that our
approach predicts a finite-temperature metal-insulator transition
between a pseudogap phase and a Mott-Hubbard insulator as 
the strength of the Coulomb interaction increases: at a critical value $U_c$,
the density of states at zero energy $\rho(\omega=0)$ 
vanishes and the pseudogap becomes a Mott-Hubbard gap
(Fig.~\ref{fig0diag0phases0magentique}). $U_c$ is obtained by equating
the minimum energy $\Delta_0$ of a HF fermion to the maximum energy
of a Schwinger boson $\sqrt{ m^2 + c^2 \Lambda^2}$. For $T \rightarrow 0$
the result is $U_c \simeq 4.25 t$.
It should be noted that the NL$\sigma$M, which is a low-energy theory,
does not allow us to describe accurately the high-energy Schwinger
bosons (with  $|\mathbf{q} | \sim \Lambda$) and in turn the low-energy
fermion excitations. In particular, the critical value of $U$ calculated 
above 
depends on the cut-off procedure used in the
NL$\sigma$M. Note also that we do
not know at which temperature and how the metal-insulator transition ends.

Fig.~\ref{fig0dos1} shows the lines $\rho(\omega=0)={\rm const}$ in
the $(U,T)$ plane. Our results are in (semi-quantitative) agreement
with the numerical calculation of Moukouri {\it et
  al.}\cite{Moukouri01} Using the criterion
$\rho(\omega=0)<10^{-2}/(2t)$ to identify the Mott-insulating phase,
these authors concluded that the system is always insulating at low
(but finite) temperature even in the weak-coupling limit, which seems
to invalidate the Slater scenario as the mechanism for the
metal-insulator transition (which requires $T_{\rm MIT}=T_{\rm
  N}=0$). Our approach shows that the results of
Ref.~\onlinecite{Moukouri01} are not in contradiction with a Slater
scenario at weak coupling, but merely reflect the exponential
suppression of the density of states due to the presence of a
pseudogap. A similar conclusion was reached in
Ref.~\onlinecite{Kyung03}.

\section{Attractive Hubbard model} 
\label{sec0hubbard0attractif}

In this section, we show that the results obtained in the previous
sections translate directly to the attractive Hubbard model. The
latter is defined by the Hamiltonian
\begin{equation}
H = - \sum_{{\bf r},\sigma} c^\dagger_{{\bf r}\sigma} (\hat t+\mu) 
c_{{\bf r}\sigma} - U \sum_{\bf r} c^\dagger_{{\bf r}\uparrow} 
c_{{\bf r}\uparrow}  c^\dagger_{{\bf r}\downarrow} 
c_{{\bf r}\downarrow} ,
\end{equation}
where $-U$ ($U\geq 0$) is the on-site attraction. $\mu=-U/2$ at
half-filling. 

Under the particle-hole transformation \cite{Micnas90}
\begin{equation}
c_{{\bf r}\downarrow} \to (-1)^{\bf r} c^\dagger_{{\bf
    r}\downarrow}, \:\:\:\:\:\:\:\:\:\:
c^\dagger_{{\bf r}\downarrow} \to (-1)^{\bf r} c_{{\bf r}\downarrow}, 
\label{phtr}
\end{equation}
the Hamiltonian becomes (up to a constant term)
\begin{eqnarray}
H &=& - \sum_{{\bf r},\sigma} c^\dagger_{{\bf r}\sigma} (\hat t+U/2) 
c_{{\bf r}\sigma} + U \sum_{\bf r} c^\dagger_{{\bf r}\uparrow} 
c_{{\bf r}\uparrow}  c^\dagger_{{\bf r}\downarrow} 
c_{{\bf r}\downarrow}
\nonumber \\ 
&& - (\mu+U/2) \sum_{\bf r}( c^\dagger_{{\bf r}\uparrow} 
c_{{\bf r}\uparrow}  - c^\dagger_{{\bf r}\downarrow} 
c_{{\bf r}\downarrow} ), 
\label{Htrans}
\end{eqnarray}
and the charge-density and pairing operators transform as
\begin{eqnarray}
\rho_{\bf r} &=& \sum_\sigma c^\dagger_{{\bf r}\sigma} c_{{\bf r}\sigma}
\to 2S^z_{\bf r}+1 , \\
\Delta_{\bf r} &=& c_{{\bf r}\downarrow} c_{{\bf r}\uparrow}
\to (-1)^{\bf r} S^-_{\bf r} , \\
\Delta^\dagger_{\bf r} &=& c^\dagger_{{\bf r}\uparrow} c^\dagger_{{\bf
    r}\downarrow}  \to (-1)^{\bf r} S^+_{\bf r} , 
\end{eqnarray}
where ${\bf S}_{\bf r}=c^\dagger_{\bf r}\mbox{\boldmath$\sigma$}
  c_{\bf r}/2$ and $S^{\pm}_{\bf r}=S^x_{\bf r}\pm iS^y_{\bf r}$.  
The transformed Hamiltonian (\ref{Htrans}) corresponds to the
repulsive half-filled Hubbard model with a uniform magnetic field
$\mu+U/2$ along the $z$ axis coupled to the fermion spins. At
  half-filling ($\mu=-U/2$), the latter 
vanishes and the Hamiltonian (\ref{Htrans}) reduces to the one studied
in the previous sections. Thus, in the attractive model,
$\mathbf{q}=\mbox{\boldmath$\pi$}$ charge and ${\bf
q}=0$ pairing fluctuations combine to form an order parameter with
SO(3) symmetry. Away from half-filling, the degeneracy between charge
and pairing fluctuations is lifted (by the uniform magnetic field
$\mu+U/2$ in the repulsive model), and the (superconducting) order
parameter exhibits SO(2) symmetry at low temperature. As a result,
there is a Berezinskii-Kosterlitz-Thouless phase
transition to a superconducting state at a finite temperature $T_{\rm
BKT}$. \cite{Berezinskii70,Kosterlitz73,Allen99} 

\begin{figure}
	\epsfig{file=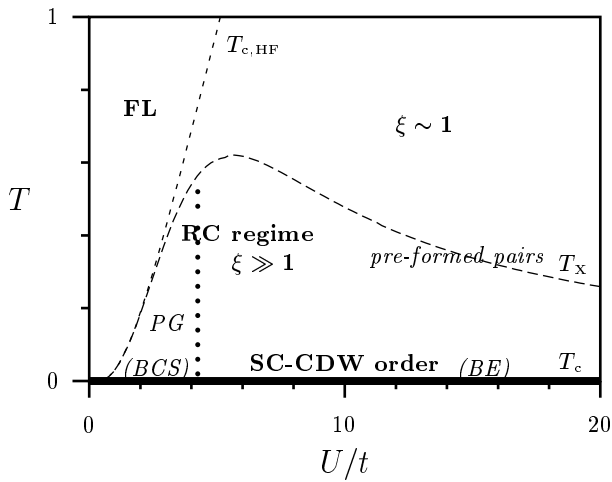}
	\caption{Phase diagram of the 2D half-filled Hubbard model
		with an attractive interaction $-U$ ($U\geq 0$).
		$T \gtrsim T_{c}^{\mathrm{HF}}$: Fermi liquid (FL) phase;
		$T_\mathrm{X} \lesssim T \lesssim
		T_{c}^{\mathrm{HF}}$: preformed pairs
		with no superfluid or
		charge-density-wave order ($\xi \sim 1$);
		$T \lesssim T_\mathrm{X}$: renormalized classical (RC) regime
		($\xi \gg 1$);
		$T=0$: superconducting (SC) and charge-density-wave (CDW)
		long-range orders ($U \ll 4t$: BCS limit, $U \gg 4t$:
		Bose-Einstein (BE) limit). 
		The dotted line is obtained from the vanishing of the tunneling
                DOS $\rho(\omega=0)$ at zero energy. 
                All lines, except $T_c=0$
		(thick solid line), are crossover lines. 
		} 
\label{fig0supra}
\end{figure}

In the following, we consider only the half-filled case where the
attractive model maps onto the repulsive model studied in the present
work. Since the 
Green's function and the spectral function are invariant under the
particle-hole transformation (\ref{phtr}), we can directly apply
the results obtained in the previous sections. The phase diagram is
shown in Fig.~\ref{fig0supra}. The crossover lines are the same as in
Fig.~\ref{fig0diag0phases0magentique}, but their physical meaning is
different. Below the
HF transition temperature $T_c^{\rm HF}$, the SO(3) order parameter
$(\rho_{\mathbf{q}=\mbox{\boldmath$\pi$}},\Delta_{\mathbf{q}=0})$ 
acquires a finite amplitude $\Delta_0$. This corresponds to the appearance of
bound particle-hole and particle-particle pairs with a size $\xi_0\sim
t/\Delta_0$. Below $T_X$, directional correlations of the order
parameter $(\rho_{\mathbf{q}=\mbox{\boldmath$\pi$}},\Delta_{\mathbf{q}=0})$
start to grow exponentially (renormalized classical regime) and
eventually long-range order sets in at the $T_c=0$ phase 
transition. Because of the SO(3) symmetry, the ground state can have
any combination of superconducting and charge-density-wave long-range
orders. As $U$ increases, the ground state smoothly evolve from the
BCS to the Bose-Einstein limits. 
In the weak-coupling limit ($U\ll 4t$), there is a pseudogap
regime at finite temperature due to the directional fluctuations of the SO(3)
order parameter. In the strong-coupling limit ($U\gg 4t$), between
$T_c^{\rm HF}$ and $T_X$, there is a regime of 
preformed (local) particle-particle pairs with no
superfluid or charge-density-wave short-range order ($\xi\sim
1$). Only below $T_X$ do these 
bosonic pairs begin to develop short-range order. At $T=0$, the
particle-particle pairs Bose condense and/or
localize, thus giving rise to superfluid and/or 
charge-density-wave long-range orders. The dotted line in
Fig.~\ref{fig0supra} is obtained from the vanishing of the tunneling
DOS $\rho(\omega=0)$ at zero energy.

\section{Summary and conclusion} 
\label{sec0conclusion}

We have presented an approach to the 2D half-filled Hubbard model
which describes both collective spin fluctuations
and single-particle properties for any value of the Coulomb repulsion
$U$. It is valid below a crossover temperature $T_X$ where amplitude
fluctuations of the AF order parameter are frozen out and AF
short-range order starts to grow exponentially (renormalized classical
regime). 

The magnetic phase diagram is obtained from a NL$\sigma$M that is
derived from the Hubbard model. The parameters of the NL$\sigma$M, the
bare spin stiffness $\rho^0_s$ and the spin-wave velocity $c$, are
expressed in terms of the mean-value of the kinetic energy and
current-current correlation functions in the HF state. The
model is solved by a saddle-point approximation within the CP$^1$
representation where the N\'eel field is represented by two Schwinger
bosons. Bose-Einstein condensation of the Schwinger bosons at zero
temperature signals the appearance of AF long-range order. At 
finite temperature (below $T_X$), the system is in a renormalized
classical regime where the AF correlation length $\xi$ is
exponentially large. The single-particle properties are obtained by
writing the fermion field in terms of a Schwinger boson and a
pseudo-fermion whose spin is quantized along the (fluctuating) N\'eel
field. This decomposition allows us to approximate the fermion Green's
function by the product (in real space) of the Schwinger boson
propagator (which is obtained from the NL$\sigma$M) and the
HF fermionic propagator. 

Our results are summarized in Fig.~\ref{fig0diag0phases0magentique}, 
which shows the phase diagram of
the 2D half-filled Hubbard model, and 
Figs.~\ref{fig0fc0spec0tnulle0ufaible}-\ref{fig0dos0tfinie0ufort}. 
At weak coupling and
zero temperature, our theory clearly describes a Slater antiferromagnet
with an exponentially small AF gap, well-defined Bogoliubov QP's
carrying most of the spectral weight, and an incoherent excitation
background at higher energy. As $U$ increases, the Slater
antiferromagnet progressively evolves into a Mott-Heisenberg
antiferromagnet with an AF gap of order $U$ and a significant
fraction of spectral weight transferred from the Bogoliubov QP's to the
incoherent excitation background. At finite temperature, the
Bogoliubov QP's disappear and only  incoherent excitations
survive. Nevertheless, precursors of the zero-temperature Bogoliubov
QP's show up as sharp peaks in the fermion spectral function, with a
width of order $T$. The presence of thermal spin fluctuations gives rise to
fermionic states below the zero-temperature AF gap. At weak coupling,
the latter is completely filled and replaced by a pseudogap. The DOS
$\rho(\omega)$ remains however exponentially small at low
energy. At strong coupling and finite temperature ($0<T\lesssim T_X
\sim J$), the system is a paramagnetic Mott-Hubbard insulator in a 
renormalized classical regime 
of spin fluctuations. At higher temperature, $T_X\sim J \lesssim T\lesssim
T_{\rm N}^{\rm HF}$, the system is characterized by the presence of
preformed local moments without AF short-range order. Thus our theory
predicts a metal-insulator transition at finite temperature between a
pseudogap phase at weak coupling and a Mott-Hubbard insulator at
strong coupling. For the 3D Hubbard
model, we expect a similar phase diagram, but 
with $T_X$ replaced by a true transition line $T_c$ between a
paramagnetic phase and an AF phase. The weak coupling
pseudogap phase therefore appears as a consequence of the low
dimensionality of the system and the high symmetry (i.e. SO(3)) of the
AF order parameter.  

At half-filling the attractive and repulsive Hubbard models can be
mapped onto one another by a canonical transformation so that our
results also apply to the attractive case. AF fluctuations in the repulsive
model correspond to $\mathbf{q}=\mbox{\boldmath$\pi$}$
charge and $\mathbf{q}=0$ pairing fluctuations in the
attractive model. The corresponding phase diagram is discussed in 
Sec.~\ref{sec0hubbard0attractif} (see Fig.~\ref{fig0supra}).

Besides its validity both at weak and strong coupling, our approach
differs from previous weak-coupling
theories\cite{Lee73,Sadovskii74,Sadovskii79,Tchernyshyov99,Bartosch99,Millis00,Schmalian98,Schmalian99,Kampf89,Kampf90,Monthoux93,Altshuler95,Vilk97a,Abanov00,Posazhennikova02}
of the pseudogap phase in
two respects. First, it takes spin fluctuations into account within a
highly non-Gaussian theory (the NL$\sigma$M) and is valid at low
temperature ($0\leq T\ll T_X$). On the contrary, most of the other
approaches assume Gaussian spin fluctuations so that their range of
validity is restricted to $T\sim T_X$. Second, our NL$\sigma$M
approach is an expansion about the AF ordered state which is a valid
starting point in presence of AF short-range order. When calculating
fermion propagators, we have to consider HF pseudo-fermions
interacting with Schwinger bosons whose dynamics is determined by the
NL$\sigma$M. Since the HF pseudo-fermions are gapped, we
expect a perturbative expansion in the pseudo-fermion-boson
interaction to be well-behaved. Our results were obtained to lowest
order where the fermion Green's function is given by the product (in
real space) of the HF fermionic propagator and the  Schwinger boson
propagator (which is obtained from the NL$\sigma$M). This should be
contrasted with perturbative treatments applied to free fermions
interacting with soft collective fluctuations where no small expansion
parameter is available. 

Our NL$\sigma$M approach is reminiscent of slave-fermion theories
\cite{Jayaprakash89,Yoshioka89,Auerbach91}
where the fermion is written as the product of a spinless
pseudo-fermion and a Schwinger boson carrying the spin degrees of
freedom. Slave-fermion theories apply to the $t$-$J$ model where the
Hilbert space is truncated by forbidding double occupancy of the
lattice sites. In our work, the pseudo-fermion also carries a spin,
which is a necessary condition to describe both the weak and strong
coupling regimes. 

Our approach bears also some analogies with the work of Gusynin {\it
et al.}\cite{Loktev01,Gusynin00,Gusynin02} on 2D fermion systems
with an attractive interaction. These authors use a ``modulus-phase''
representation for the SO(2) superconducting order parameter which is
analog to our ``amplitude-direction'' representation of the SO(3)
AF order parameter. At low temperature, the phase of the
superconducting order parameter is governed by a SO(2) sigma
model. The fermion Green's function is calculated both above and below the
Berezinskii-Kosterlitz-Thouless phase transition $T_{\rm BKT}$ by writing
the fermion field as the product of a pseudo-fermion and a bosonic
field which is related to the phase of the order parameter. As in our
work, a simple decoupling procedure between pseudo-fermions and bosons
is used. A pseudogap phase is found both above and below $T_{\rm
BKT}$. Gusynin {\it et al.} also point out the necessity to
perform a fully quantum-mechanical calculation to describe the pseudogap
phase. \cite{Gusynin02} The main difference with our work comes from
the SO(2) symmetry of the order parameter which leads to a
finite-temperature Berezinskii-Kosterlitz-Thouless phase
transition. 

Let us now  mention some limitations of our approach. (i) The feedback of
spin fluctuations on (pseudo)-fermions is not fully taken into
account. As a result, we miss important effects, like the
renormalization of the zero-temperature HF gap $\Delta_0$ by quantum
spin fluctuations. (ii) The crossover temperature $T_X$, which is
identified to the HF transition temperature $T_{\rm N}^{\rm HF}$ at
weak coupling, is overestimated. Due to Kanamori screening effects,
$T_X$ should be smaller than $T_{\rm N}^{\rm
  HF}$. \cite{Vilk97,Kyung03} (iii) The NL$\sigma$M approach 
is restricted to low temperature ($T\ll T_X$). In particular, it does
not give access to the crossover regime between the Fermi liquid and
the pseudogap phase at weak coupling. This regime is characterized, as
the temperature decreases, by the suppression of Landau's QP's. (iv)
At finite temperature, we predict a metal-insulator transition between a
pseudogap phase and a Mott-Hubbard insulator. However, being a
low-energy theory, the NL$\sigma$M does
not allow to study the finite-temperature metal-insulator transition
in detail (see Sec.~\ref{sec0fc0spec}). 

But the main shortcoming of our approach is that it does not
distinguish between Bogoliubov and Mott-Hubbard bands. We find a 
single energy scale ($\Delta_0$) in the density of states
$\rho(\omega)$ and the spectral function ${\cal A}(\mathbf{k},\omega)$. On
physical grounds, we expect instead two energy scales, namely
$\Delta_0$ and $U/2$, corresponding to Bogoliubov bands (or precursors
thereof at finite temperature) and Mott-Hubbard bands,
respectively.\cite{Vilk97} In the weak coupling limit, 
$\Delta_0$ depends crucially on the nesting properties of the Fermi
surface (Slater antiferromagnetism). On the other hand, the energy
scale $U/2$ has a purely local origin, which is independent of the
Fermi surface geometry, and is associated with the Mott-Hubbard
localization. A proper description of the Mott-Hubbard localization
would require to treat the charge fluctuations beyond the HF
approximation for the $\Delta_c$ field (Sec.~\ref{sec0derivation}). In the
strong-coupling limit, charge fluctuations are frozen out. 
This is the reason why the HF saddle point for the amplitude 
fields $\Delta_c$ and $\Delta_s$ provides an accurate description
of the local moments (whose direction is given by the $\bf\Omega_{\bf
r}$ field) which form in the strong coupling limit.\cite{note5}
Note that for $U\gg 4t$,  $\Delta_0\to U/2$ so that the system is
characterized by a single energy scale. At intermediate coupling
($U\sim 8t$), a four-peak structure corresponding to the simultaneous
presence of Bogoliubov and Mott-Hubbard bands has been observed in 
numerical simulations \cite{Preuss95,Moreo95} and analytical
studies \cite{Vilk97,Matsumoto97} of the Hubbard model. Although it
misses some aspects of the Mott-Hubbard localization, in particular at
intermediate coupling, we believe that our theory captures the main
features of the physics of the 2D half-filled Hubbard model. 

There are several directions in which this work could be further
developed. The most obvious one is to consider situations where
antiferromagnetism is frustrated due to either a non-bipartite
lattice or a finite next-neighbor hopping amplitude. Doping would also
induce magnetic frustration. This opens up the possibility to
stabilize more exotic magnetic orders (e.g. a non-collinear order),
and/or to reach the quantum disordered and quantum critical regimes of
the NL$\sigma$M (Fig.~\ref{fig0msnl}) and study the corresponding
fermion spectral functions. 

\acknowledgments

We would like to thank A.M.-S. Tremblay for numerous discussions
and for pointing out Ref.~\onlinecite{Loktev01}.


\appendix

\section{ HF current-current correlation function}
\label{app0correltateur}

In this appendix we calculate the static uniform current-current
correlation function
\begin{eqnarray}
		\Pi_{\mu \mu'}^{\nu \nu'}
	=
		\left<
			j_{\mu}^{\nu}( \mathbf{0} , 0 )
			j_{\mu'}^{\nu'}( \mathbf{0} , 0 )
		\right>_{ \mathrm{HF} } .
	\label{eq0def0pi0zero}
\end{eqnarray}
From the definition of the
current $j_\mu^\nu$ [Eqs.~(\ref{eq0j0temps}-\ref{eq0j0espace})], 
we see that its zero-frequency
zero-momentum Fourier transform involved in Eq.~(\ref{eq0def0pi0zero}) is given
by
\begin{eqnarray}
		j_{\mu}^{\nu}( \mathbf{0} , 0 )
	&=&
		\frac{1}{ \sqrt{\beta \mathcal{N}} }
		\sum_{ \mathbf{k} , \omega }
		v_\mu ( \mathbf{k} )
		\Phi_{ \mathbf{k} \omega }^{\dagger}
		\sigma_\nu
		\Phi_{ \mathbf{k} \omega }
	\label{eq0j} ,
\end{eqnarray}
where
\begin{eqnarray}
		v_0 ( \mathbf{k} )
	&=&
		1
	\label{eq0a0zero} , \\
		v_\mu ( \mathbf{k} )
	&=&
		2 t \sin ( k_\mu )
	, \qquad
		\mu = x , y
	\label{eq0a0mu} .	
\end{eqnarray}
Using the Wick's theorem to evaluate HF averages of $\Phi$ fields,
we can express $\Pi_{\mu \mu'}^{\nu \nu'}$ as
\begin{widetext}
\begin{eqnarray}
		\Pi_{\mu \mu'}^{\nu \nu'}
	&=&
		- \frac{1}{\beta \mathcal{N} }
		\sum_{ \mathbf{k} , \omega \atop \mathbf{k}' , \omega' }
		v_\mu ( \mathbf{k} ) v_{\mu'} ( \mathbf{k}' )
		\mathrm{Tr}
		\left[
			\sigma_\nu
			\left<
				\Phi_{ \mathbf{k} \omega }
				\Phi_{ \mathbf{k}' \omega' }^{\dagger} 
			\right>
			\sigma_{\nu'}
			\left<
				\Phi_{ \mathbf{k}' \omega' }
				\Phi_{ \mathbf{k} \omega }^{\dagger} 
			\right>
		\right] ,
	\label{eq0pi0zero}
\end{eqnarray}
where Tr denotes the trace with respect to the spin indices. Writing the HF
propagator [Eq.~(\ref{eq0green0hf})] as 
\begin{eqnarray}	
		-
		\left<
			\Phi_{ \mathbf{k} \omega }
			\Phi_{ \mathbf{k}' \omega' }^{\dagger} 
		\right>
	&=&
		\delta_{ \omega , \omega' } 
		\left[	
			\delta_{ \mathbf{k} , \mathbf{k}' }
			G( \mathbf{k} , \omega )
		+
			\delta_{ \mathbf{k} , \mathbf{k}' + \boldsymbol{\pi} }		
			\sigma_3
			F( \mathbf{k} , \omega )
		\right]
	\label{eq0prop0hf0app} , \\
		G( \mathbf{k} , \omega )
	&=&	
		\frac	{ - i \omega - \epsilon_{ \mathbf{k} } }
			{ \omega^2 + E_{ \mathbf{k} }^2 }
	, \qquad
		F( \mathbf{k} , \omega )
	=
		\frac	{ \Delta_0 }
			{ \omega^2 + E_{ \mathbf{k} }^2 }
	\label{eq0prop0fg} ,
\end{eqnarray}	 
and using 
$ \mathrm{Tr} ( \sigma_\nu \sigma_{\nu'} ) = 2 \delta_{ \nu , \nu' } $,
$ \mathrm{Tr} ( \sigma_3 \sigma_\nu \sigma_3 \sigma_{\nu'} )= 2 \delta_{
  \nu , \nu' } (2\delta_{\nu,3}-1)$ 
and
$F( \mathbf{k} + \boldsymbol{\pi} , \omega ) = F( \mathbf{k} , \omega )$,
we obtain
\begin{eqnarray}
		\Pi_{\mu \mu'}^{\nu \nu'}
	&=&
		- \frac{ 2 \delta_{ \nu , \nu' } }{\beta \mathcal{N}}
		\sum_{ \mathbf{k} , \mathbf{k}' , \omega }
		v_\mu ( \mathbf{k} ) v_{\mu'} ( \mathbf{k}' )
		\left[
			\delta_{ \mathbf{k} , \mathbf{k}' }
			G( \mathbf{k} , \omega )^2
		+
			\delta_{ \mathbf{k} , \mathbf{k}' + \boldsymbol{\pi} }
			( 2 \delta_{\nu,3} - 1 )
			F( \mathbf{k} , \omega )^2
		\right]	.	
	\label{eq0pi0gf}
\end{eqnarray}	 
$\Pi_{\mu \mu'}^{\nu \nu'}$ is thus diagonal in $\nu$ and $\nu'$. One can show
that it is also diagonal in $\mu$ and $\mu'$. Indeed, whenever these
two indices are different, the rhs of  (\ref{eq0pi0gf}) is odd in
$k_x$ or $k_y$ and vanishes after wave-vector summation. Furthermore, 
$v_0 ( \mathbf{k} + \boldsymbol{\pi} ) = v_0 ( \mathbf{k} )$
and
$v_\mu ( \mathbf{k} + \boldsymbol{\pi} ) = - v_\mu ( \mathbf{k} )$
for $\mu = x,y$, so that
\begin{eqnarray}
		\Pi_{\mu \mu'}^{\nu \nu'}
	&=&
		- \frac	{ 2 \delta_{ \nu , \nu' } \delta_{ \mu , \mu' }}
			{\beta \mathcal{N}}
		\sum_{ \mathbf{k} , \omega }
		v_\mu ( \mathbf{k} )^2
		\left[
			G( \mathbf{k} , \omega )^2
		+
			( 2 \delta_{\mu,0} - 1 )
			( 2 \delta_{\nu,3} - 1 )
			F( \mathbf{k} , \omega )^2
		\right]	.
\end{eqnarray}	
\end{widetext}
For $T\ll T_{\rm N}^{\rm HF}$, one can perform the 
Matsubara frequency summation in the zero-temperature limit. This gives
\begin{eqnarray}
		-
		\frac{1}{\beta}
		\sum_{ \omega }
		G( \mathbf{k} , \omega )^2
	&=&
		\frac{1}{\beta}
		\sum_{ \omega }
		F( \mathbf{k} , \omega )^2
	\nonumber \\
	&=&
		\frac{ \Delta_0^2 }{4 E^3_{ \mathbf{k} } } .	
	\label{eq0somme0f}
\end{eqnarray}	 
The only non-vanishing correlator functions are therefore
\begin{eqnarray}
		\Pi_{0 0}^{1 1}
	&=&
		\Pi_{0 0}^{2 2} \nonumber \\ 
	&=&
		\int_{ \mathbf{k} }
		\frac{\Delta_0^2}{ E_{ \mathbf{k} }^3 }
	\label{eq0pi0temps} , \\
		\Pi_{x x}^{3 3}
	&=&
		\Pi_{y y}^{3 3} \nonumber \\
	&=&
		4 \Delta_0^2 t^2
		\int_{ \mathbf{k} }
		\frac{\sin^2 k_x}{ E_{ \mathbf{k} }^3 } .
\end{eqnarray}

\section{$\mathrm{SU}(2)$ gauge field} 
\label{app0jauge}

In this appendix we give a proof of
Eqs.~(\ref{eq0a0et0n01}-\ref{eq0a0et0n02}), relating the N\'eel and canting
fields $\mathbf{n}_\mathbf{r}$ and $\mathbf{L}_\mathbf{r}$ to the gauge field
$A_{\mu \mathbf{r}}^{\nu}$ and the rotated canting field
$\mathbf{l}_\mathbf{r}$. Let us recall the definition of the gauge field:
\begin{equation}
		A_{ \mu \mathbf{r} } 
	= 
		i R_\mathbf{r}^{\dagger} \partial_\mu R_\mathbf{r}, 
                \qquad \mu = t,x,y. 
	\label{eq0def0a}
\end{equation}
The index $t$ stands for real-time derivation. Imaginary-time
results are obtained using $\tau = i t$. The
$\mathrm{SU}(2)/\mathrm{U}(1)$ rotation  
matrix $R_\mathbf{r}$ is defined, up to a $\mathrm{U}(1)$ gauge transformation 
$R_\mathbf{r} \rightarrow R_\mathbf{r} e^{ i \alpha_\mathbf{r} \sigma_3 }$,
by 
\begin{equation}
		\boldsymbol{\sigma} \cdot \mathbf{n}_\mathbf{r}
	=
		R_\mathbf{r} \sigma_3 R_\mathbf{r}^{\dagger} ,
	\label{eq0def0r}
\end{equation}
which means that the SO(3) element $\mathcal{R}_\mathbf{r}$ associated to $R_\mathbf{r}$
maps $\mathbf{u}_z$ onto $\mathbf{n}_\mathbf{r}$.
The gauge field $A_{ \mu \mathbf{r} }^{\nu}$ is a  zero-trace
Hermitian matrix which can be decomposed on Pauli
matrices $\sigma_\nu$:
\begin{eqnarray}
		A_{\mu \mathbf{r} } 
	&=& 
		\sum_{\nu = 1,2,3}
		A_{\mu \mathbf{r}}^\nu \sigma_\nu \nonumber \\ 
	&=& 
		\mathbf{A}_{\mu\mathbf{r}} \cdot \boldsymbol{\sigma}
	\label{eq0a0decomp}
\end{eqnarray}
where the bold notation denotes the three-component vector $( A_\mu^1
, A_\mu^2 , A_\mu^3 )$. 

The main result of this appendix is the following general form for the
$\mathbf{A}_{\mu \mathbf{r}}$ field:
\begin{eqnarray}
		\mathbf{A}_{\mu \mathbf{r}}
	&=&
		\mathcal{R}_\mathbf{r}^{-1}
		\left(
			\frac{1}{2} \mathbf{n}_\mathbf{r} 
			\wedge \partial_\mu \mathbf{n}_\mathbf{r}
			+
			\kappa_{\mu \mathbf{r}} \mathbf{n}_\mathbf{r}
		\right)
	\label{eq0fond01} \\
	&=&
		\frac{1}{2} \mathbf{u}_z \wedge 
		\mathcal{R}_\mathbf{r}^{-1}( \partial_\mu \mathbf{n}_\mathbf{r} )
		+
		\kappa_{\mu \mathbf{r}} \mathbf{u}_z
	. \label{eq0fond02}
\end{eqnarray}
$\kappa_{\mu \mathbf{r}}$ is some function of position and time, fixed by the choice of a 
gauge. Notice however, that it cannot be any function, since it appears
in the expression of the gauge-field density tensor, which must be zero.

Eqs.~(\ref{eq0a0et0n01}-\ref{eq0a0et0n02}) follow quite easily. First, 
we have
\begin{eqnarray}
		\sum_{\nu = 1,2} { A_{\mu \mathbf{r}}^\nu }^2
	&=&
		\frac{1}{4}
		\left[ 
			\mathcal{R}_\mathbf{r}^{-1}
			( 
				\mathbf{n}_\mathbf{r} 
				\wedge \partial_\mu 
				\mathbf{n}_\mathbf{r} 
			) 
		\right|^2
	\nonumber \\
	&=&
		\frac{1}{4}
		( 
			\mathbf{n}_\mathbf{r} 
			\wedge \partial_\mu 
			\mathbf{n}_\mathbf{r} 
		)^2 
	\nonumber \\
	&=&
		\frac{1}{4}
		( 
			\partial_\mu \mathbf{n}_\mathbf{r} 
		)^2 
	\label{eq0calcul0aa} .
\end{eqnarray}
Using $\partial_t = i \partial_\tau$ we obtain Eq.~(\ref{eq0a0et0n01}). Second, recalling
that the rotated canting vector
$\mathbf{l}_\mathbf{r} = \mathcal{R}_\mathbf{r}^{-1} \mathbf{L}_\mathbf{r}$ has no 
component along $\mathbf{u}_z$, we can write
\begin{eqnarray}
		\sum_{\nu = 1,2} A_{\mu \mathbf{r}}^\nu l_{\mathbf{r}}^{\nu}
	&=&
		\mathbf{A}_{\mu \mathbf{r}}
		\cdot
		\mathbf{l}_\mathbf{r}
	\nonumber \\
	&=&
		\frac{1}{2}
		\mathcal{R}_\mathbf{r}^{-1}
			( 
				\mathbf{n}_\mathbf{r} 
				\wedge \partial_\mu 
				\mathbf{n}_\mathbf{r} 
			) 
		\cdot
		\mathcal{R}_\mathbf{r}^{-1}
			(
				\mathbf{L}_\mathbf{r}
			)
	\nonumber \\
	&=&
		\frac{1}{2}
		( 
			\mathbf{n}_\mathbf{r} 
			\wedge \partial_\mu 
			\mathbf{n}_\mathbf{r} 
		) 
		\cdot
		\mathbf{L}_\mathbf{r} ,
	\label{eq0calcul0al}
\end{eqnarray}	 
hence Eq.~(\ref{eq0a0et0n02}). 

We now give a derivation of Eq.~(\ref{eq0fond02}).
The first step is to differentiate Eq.~(\ref{eq0def0r}). Derivatives
of  the rotation matrix are calculated  using Eq.~(\ref{eq0def0a}) and
the identity 
$ \partial_\mu R_\mathbf{r}^{\dagger} = - R_\mathbf{r}^{\dagger} 
( \partial_\mu R_\mathbf{r} ) R_\mathbf{r}^{\dagger}$
which results from the unitarity of $R_\mathbf{r}$.
We obtain
\begin{eqnarray}
		\boldsymbol{\sigma} \cdot \partial_\mu \mathbf{n}_\mathbf{r}
	&=&
		-i R_\mathbf{r}	A_{\mu \mathbf{r}} \sigma_3
	R_\mathbf{r}^{\dagger}  
		+i R_\mathbf{r} \sigma_3 A_{\mu \mathbf{r}}
	R_\mathbf{r}^{\dagger}  
   \nonumber \\
	&=&
		-i R_\mathbf{r}	A_{\mu \mathbf{r}} R_\mathbf{r}^{\dagger} 
		 R_\mathbf{r}\sigma_3 R_\mathbf{r}^{\dagger} 
		+
		i R_\mathbf{r} \sigma_3 R_\mathbf{r}^{\dagger} 
		R_\mathbf{r} A_{\mu \mathbf{r}} R_\mathbf{r}^{\dagger} 
	\nonumber .	\nonumber \\ && \label{eq0differentiation}
\end{eqnarray}	 
Let us define a new field
\begin{eqnarray}
		\tilde{A}_{\mu \mathbf{r}}  
        &=& 
		R_\mathbf{r} A_{\mu \mathbf{r}} R_\mathbf{r}^{\dagger}
		\nonumber \\ 
	&=&
		\mathbf{\tilde{A}}_{\mu \mathbf{r}} \cdot
		\boldsymbol{\sigma} .
\end{eqnarray}
Using $\tilde{A}_{\mu \mathbf{r}}$ and Eq.~(\ref{eq0def0r}) we can rewrite
Eq.~(\ref{eq0differentiation}) as
\begin{eqnarray}
		\boldsymbol{\sigma} \cdot \partial_\mu \mathbf{n}_\mathbf{r}
	&=&
		-i
		\big[
			\tilde{A}_{\mu \mathbf{r}} ,
			\boldsymbol{\sigma} \cdot \mathbf{n}_\mathbf{r} 
		\big]
	\nonumber \\
	&=&
		-i
		\big[
			\boldsymbol{\sigma} \cdot \mathbf{\tilde{A}}_{\mu \mathbf{r}} ,
			\boldsymbol{\sigma} \cdot \mathbf{n}_\mathbf{r} 
		\big]
	\nonumber \\
	&=&
		2
		\boldsymbol{\sigma}
		\cdot
		\left(
			\mathbf{\tilde{A}}_{\mu \mathbf{r}} 
			\wedge \mathbf{n}_\mathbf{r}
		\right)
	. \label{eq0differentiation02}
\end{eqnarray}	 
We have used the identity
\begin{eqnarray}
		[
			\boldsymbol{\sigma} \cdot \mathbf{u} 
			,
			\boldsymbol{\sigma} \cdot \mathbf{v} 
		]
	=
		2 i
		\boldsymbol{\sigma} \cdot 
		( \mathbf{u} \wedge \mathbf{v} )	
	\label{eq0crochets} .
\end{eqnarray}
Identifying the coefficients of $\boldsymbol{\sigma}$ in
Eq.~(\ref{eq0differentiation02}) and  vector-multiplying by
$\mathbf{n}_\mathbf{r}$ we arrive at 
\begin{eqnarray}
		\mathbf{\tilde{A}}_{\mu \mathbf{r}}
	=
		\frac{1}{2}
		\mathbf{n}_\mathbf{r} \wedge
		\partial_\mu \mathbf{n}_\mathbf{r}
		+
		(
			\mathbf{n}_\mathbf{r} \cdot
			\mathbf{\tilde{A}}_{\mu \mathbf{r}}
		)    \mathbf{n}_\mathbf{r} .
	\label{eq0rel0a0dn}
\end{eqnarray}	 
To conclude, it is sufficient to define the last term in
Eq.~(\ref{eq0rel0a0dn}) as 
$\kappa_{\mu \mathbf{r}}$ and to remark that, owing to the definition of
$\tilde{A}_{\mu \mathbf{r}}$, we have 
$\mathbf{\tilde{A}}_{\mu \mathbf{r}} = \mathcal{R}_\mathbf{r} 
\mathbf{A}_{\mu \mathbf{r}}$.



\begin{thebibliography}{99}
\bibitem{note4} Our approach is closer to slave-fermion theories since
  the spin degrees of freedom are represented by Schwinger bosons.
\bibitem{note7} $2\Delta_0$ corresponds to the threshold energy for
  particle-hole excitations in the HF state. Since spin-wave modes
  cannot exist above   this energy, a natural choice for the
  NL$\sigma$M cutoff is $\Lambda\sim {\rm min}(1,2\Delta_0/c)$.
\bibitem{note0interp} The function used in this interpolation is
$T_X = \sqrt{ {T_{\mathrm{N}}^{\mathrm{HF}}}^2 + {T'}^2 } $.
\end{thebibliography}
\end{document}